# Mono-exponential Current Attenuation with Distance across 16 nm Thick Bacteriorhodopsin Multilayers

*Domenikos Chryssikos, Jerry A. Fereiro, Jonathan Rojas, Sudipta Bera, Defne Tüzün, Evanthia Kounoupioti, Rui N. Pereira, Christian Pfeiffer, Ali Khoshouei, Hendrik Dietz, Mordechai Sheves, David Cahen, and Marc Tornow\**

D. Chryssikos, J. A. Fereiro, J. Rojas, D. Tüzün, E. Kounoupioti, R. N. Pereira, C. Pfeiffer, M. Tornow
School of Computation, Information, and Technology
Technical University of Munich
85748 Garching, Germany
E-mail: tornow@tum.de

D. Chryssikos, R. N. Pereira, M. Tornow
Fraunhofer Institute for Electronic Microsystems and Solid State Technologies
80686 Munich, Germany

J. A. Fereiro
School of Chemistry
Indian Institute of Science Education and Research Thiruvananthapuram
695551 Kerala, India

S. Bera, M. Sheves, D. Cahen
Department of Molecular Chemistry and Materials Science
Weizmann Institute of Science
Rehovot 7610001, Israel

A. Khoshouei, H. Dietz
School of Natural Sciences
Technical University of Munich
85748 Garching, Germany

**Abstract**

The remarkable ability of natural proteins to conduct electricity in the dry state over long distances remains largely inexplicable despite intensive research. In some cases, a (weakly) exponential length-attenuation, as in off-resonant tunneling transport, extends to thicknesses even beyond 10 nm. This report deals with such charge transport characteristics observed in self-assembled multilayers of the protein bacteriorhodopsin (bR). About 7.5 nm to 15.5 nm thick bR layers were prepared on conductive titanium nitride (TiN) substrates using aminohexylphosphonic acid and poly-diallyl-dimethylammonium electrostatic linkers. Using conical EGaIn top contacts, an intriguing, mono-exponential conductance attenuation as a function of the bR layer thickness with a small attenuation coefficient $\beta \approx 0.8$ nm$^{-1}$ is measured at zero bias. Variable-temperature measurements using evaporated Ti/Au top contacts yield effective energy barriers of about 100 meV from fitting the data to tunneling, hopping, and carrier cascade transport models. The observed temperature-dependence is assigned to the protein-electrode interfaces. The transport length and temperature dependence of the current densities are consistent with tunneling through the protein-protein and protein-electrode interfaces, respectively. Importantly, our results call for new theoretical approaches to find the microscopic mechanism behind the remarkably efficient, long-range electron transport within bR.

## 1. Introduction

There has been an emerging interest in the apparent long-range charge transport through 10-50 nm thick films of nominally insulating or poorly conductive organic material.[1–9] This refers in particular to transport indicative of coherent tunneling, in cases where the film thickness is well beyond the accepted range for such transport mechanism and where coherence remains to be proven.[3–6] Those experimental results have raised the fundamental question: what is the underlying transport mechanism of this long-range charge transport?

Several groups reported very weak distance- or length-dependent attenuation of the conductance $G$, which apparently follows an exponential law,

$$G = G_C \exp(-\beta d) \quad (1)$$

with an unusually small decay coefficient $\beta$.[1–3,10–12] Here, $d$ is the thickness of the investigated material layer (i.e., the transport length), and $G_C = G(d = 0)$ is the contact (zero-length) conductance.[13,14] In many cases, two or more characteristic length regimes were observed, featuring a transition from an already small $\beta$ (at short transport lengths) to an even smaller $\beta$ (at larger transport lengths).[5–7,10] A mechanism involving multiple individual tunneling steps, "multi-step" or "sequential tunneling", was frequently invoked to explain these observations.[3,5,6,10] Examples include transport across various organic and other related thin-films, such as copper phthalocyanine layers up to 60 nm thick, with $\beta$ values < 0.1 nm$^{-1}$,[10] layers of viologen-based oligomers (14 nm, 0.25 nm$^{-1}$),[1] oligo-porphyrin molecular wires (~5 nm, 0.4 nm$^{-1}$),[11] polypyridyl oligomer layers (8 nm, 0.19 nm$^{-1}$),[12] oligo(thiophene) layers (22 nm, 0.015 nm$^{-1}$),[2] and even Prussian blue analog nanocrystals (45 nm, 0.11 nm$^{-1}$).[3]

A particularly intriguing class of thin films with related properties are proteins. Research on monolayers of certain proteins has provided evidence in recent years that they are electronically conductive in the dry state (with only a few structural water molecules retained) and when typically embedded between electrical contacts - as an integral part of a solid-state type junction.[15–17] This manifestation of efficient charge transport in dry protein films has also sparked applied interest in prospective protein-based nanoscale solid-state devices.[18] The proteins' conductance has generally been found to attenuate slower as a function of length than that of alkanes and even conjugated organic molecular wires.[17]

In contrast to layers of organic, oligomeric molecules[2,19,20] (including poly-peptides[21]), studying the length-dependence of the conductance for proteins is challenging due to their size, structural complexity, and sensitivity to environmental conditions, such as temperature and humidity. For that reason, reports of experimentally performed length-dependent charge transport studies in specific proteins are scarce. Examples include recent publications on modified E2 proteins with layer thickness ≤ 19 nm and $\beta$ value 0.39 nm$^{-1}$,[8] on ferritin (two studies) with layer thicknesses ≤ 11 nm and $\beta$ values 0.28 nm$^{-1}$ and 0.37 nm$^{-1}$ in the sequential tunneling regime,[5–7] and on consensus tetratricopeptide repeat (CTPR) proteins with layer thickness ≤ 12 nm and $\beta$ values from 0.10 nm$^{-1}$ to 0.18 nm$^{-1}$.[22] A possible alternative approach is grafting multiple protein entities together while perturbing their structure as little as possible. This way, as protein molecules can already measure a few nanometers in diameter, a multi-stack protein film, deposited layer-by-layer, similar to alternating-charge poly-peptide films,[23] may well reach thicknesses up to a few tens of nanometers or higher. Such an approach has been reported by He et al.[24] and Chu et al.[25] for bacteriorhodopsin (multilayer thickness up to 28 nm and 480 nm, respectively), and recently by Barhom et al.[26] for Photosystem I (thickness up to 30 nm). No $\beta$ values were reported in these studies.

**Table 1**: Summary of the length and temperature-dependence of the electrical conductance in various charge transport models, at low bias.

| Mechanism | Length-dependence | Temperature-dependence |
|---|---|---|
| Tunneling | $G(d) \propto \exp(-\beta d)$ | $G(T) - G(0\ \text{K}) \propto \dfrac{\text{sech}^2\left(\frac{\phi}{2k_BT}\right)}{k_BT}$ |
| Hopping | $G(d) \propto \dfrac{1}{d}$ | $G(T) \propto \exp\left(-\dfrac{\phi}{k_BT}\right)$ |
| Variable-range hopping (Mott) | $G(d) \propto \dfrac{1}{d}$ | $G(T) \propto \exp\left[-\left(\dfrac{\phi}{k_BT}\right)^{\frac{1}{D+1}}\right]$ |
| Carrier cascade | N/A | low $T$: $G(T) - G(0\ \text{K}) \propto \exp\left(-\dfrac{\Delta_{LT}}{k_BT}\right)$<br>high $T$: $G(T) \propto \exp\left(-\dfrac{\Delta_{HT}}{k_BT}\right)$ |

Symbols: $\phi$ = energy barrier height, $k_B$ = Boltzmann constant, $D$ = number of dimensions, $\Delta_{LT}$ = energy difference either between the highest occupied molecular orbital (HOMO) and the next lower level ($E_{HOMO} - E_{HOMO-1}$) or the lowest unoccupied molecular orbital (LUMO) and the next higher level ($E_{LUMO+1} - E_{LUMO}$), $\Delta_{HT}$ = energy difference between HOMO and LUMO ($\Delta_{HT} = E_{LUMO} - E_{HOMO}$).

All length-dependent protein conductance studies published so far agree in such that one (or more) unusually small $\beta$ value(s) were derived, indicating efficient long-range transport in these systems. Various transport mechanisms have been discussed in the literature to describe long-range charge transport in proteins, including thermally activated or diffusion-assisted hopping, super-exchange-assisted tunneling, flickering resonance, and even band-like conductance as known from solids with periodic crystal structure.[16,27,28] All these different charge transport mechanisms are characterized by their (often unique) length and temperature-dependence.

[14,16,29] In this work, we consider four fundamental charge transport models to interpret our observed experimental data: tunneling, thermally activated hopping, variable-range hopping (VRH), and carrier-cascade transport. The length ($d$) and temperature ($T$) dependencies of the conductance $G$ in each model are presented in the Supporting Information (SI, Section 1) and summarized in **Table 1**.

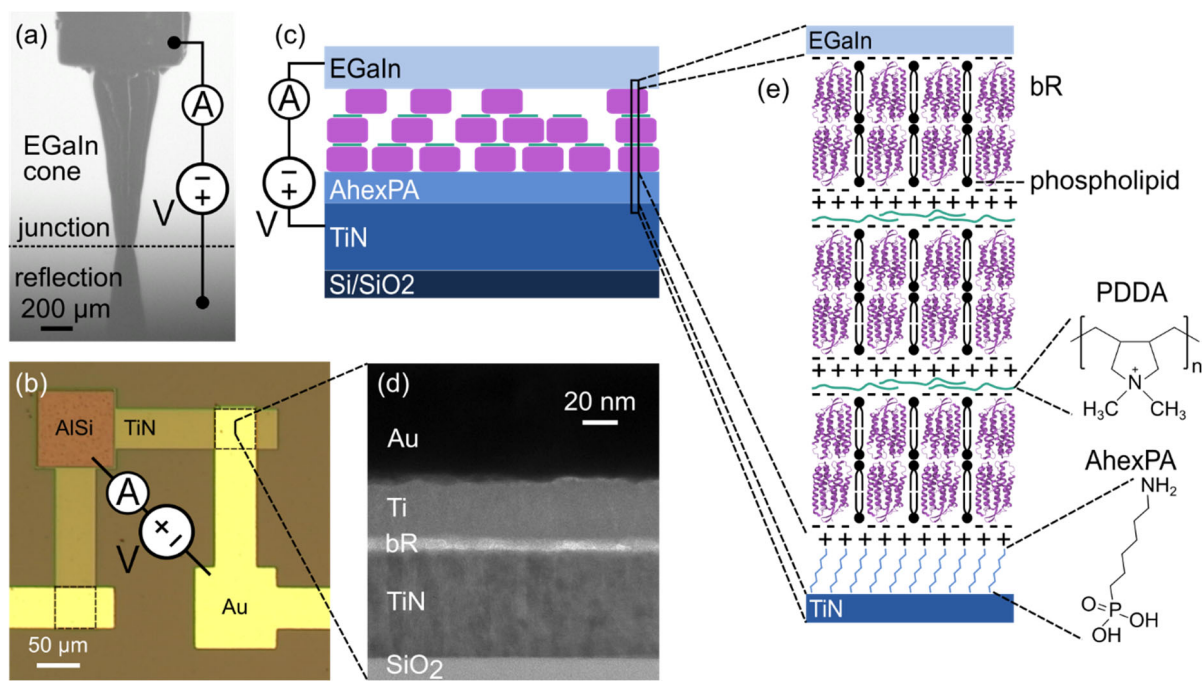

**Figure 1.** (a) Optical micrograph of a junction with eutectic gallium-indium (EGaIn) top contact and TiN bottom contact (side view), schematically showing the electrical measurement configuration. The dashed line indicates the position of the EGaIn contact with the substrate. Below the superimposed dashed line, the reflection of the EGaIn cone on the sample surface is visible. (b) Optical micrograph of two permanent junctions with evaporated Ti/Au top contact and TiN bottom contact (top view), schematically showing the electrical measurement configuration. The dashed squares indicate the junctions. (c) Schematic side view of a protein junction comprising (from bottom to top) a TiN bottom electrode, a self-assembled 6-aminohexylphosphonic acid (AhexPA) linker layer, a triple bacteriorhodopsin (bR; purple blocks) vesicle layer grafted together via poly-diallyl-dimethylammonium (PDDA; green segments) linkers, and an EGaIn electrode, including the electrical measurement configuration. (d) Cross-sectional TEM image of a junction as in (b), comprising a bR vesicle monolayer. (e) Close-up of an electrostatically bound bR vesicle trilayer as in (c), displaying electrodes, bR molecules, phospholipid molecules, and AhexPA and PDDA electrostatic linkers. The chemical formulas of the two linker molecules are shown to the right of the stack.

We focus on the membrane-spanning protein bacteriorhodopsin (bR), which functions as a light-driven proton pump in the purple membrane of the archaeon *Halobacterium Salinarum*. bR has been successfully investigated in charge transport studies in the dry state before, owing to pre-established extraction and vesicle assembly protocols,[30] the high stability of the protein at ambient conditions (also in the dry state), and its high denaturation temperature, which in dry films may exceed 140 °C.[31] Studies from the 2000's revealed that bR is relatively conductive despite its comparatively large size of roughly 5 nm,[32,33,34] with an estimated $\beta$ of 1.2 nm$^{-1}$.[35,36]

In particular, we study charge transport through stacked bR multilayers, which allows us to investigate the length-dependence of charge transport to a better accuracy than what was attainable in single bR vesicle layer studies reported before.[35,36] **Figure 1** illustrates the bR junctions used in this work. As the bottom electrode we chose conductive titanium nitride (TiN), and an organophosphonate self-assembled monolayer (SAM) to tether the first layer of bR onto the TiN bottom substrate (see Figure 1 (c)). TiN is a hard, conductive, chemically resistant ceramic that is well-established in semiconductor (CMOS) technology.[37] Owing to these properties, TiN is used here as a bottom contact material that appears to be promising for protein electronics. Organo-phosphonate SAMs are versatile, stable surface layers that can assemble on a large variety of metal and semiconductor oxides. They can serve different purposes, ranging from passivation, via acting as gate material for organic field effect transistors, up to providing the interface for bio-functionalization of sensor devices.[38] In particular, they have proven to provide an excellent platform for studying tunneling transport through alkane SAMs, revealing the anticipated exponential decay of conductance with alkane chain length in SAMs on Si/SiO$_2$[39] and on Si/AlO$_x$[40,41]. We use organophosphonates here instead of commonly used coupling agents such as silanes. Silane SAMs, while providing a continuous coverage, connect with only a fraction of the possible covalent bonds directly to the surface, because of lateral cross-linking between their headgroups.[42] In contrast, in organophosphate SAMs, each phosphonate group is bound exclusively to the substrate, which likely provides a more homogeneous charge injection from the electrode into the SAM.[43]

Bacteriorhodopsin is used in the form of vesicles after treatment of the purple membrane with the detergent octylthioglucoside (OTG). This treatment yields protein-lipid vesicles, which exhibit reduced lipid content relative to the native purple membrane, i.e., the resulting vesicles are partially delipidated. To achieve multiple bR layer stacking, we used the cationic polymer poly-diallyl-dimethylammonium (Figure 1 (e)) as an electrostatic linker between the successive bR layers, which have intrinsically negatively charged surfaces.[24,25,44] To perform electrical measurements of layer thickness dependence in a vertical configuration, we used a eutectic gallium indium (EGaIn) liquid metal electrode as top contact, as shown in Figure 1 (a). Notwithstanding its ultra-thin native oxide, EGaIn is widely used as a material to contact molecular layers in junctions and, in recent years, also layers of biomolecules.[45,46] Our results reveal a surprising *mono-exponential* attenuation of the measured current as a function of bR layer thickness up to a maximum of about 16 nm, within the accuracy range of our data. The attenuation is characterized by a single small decay constant *β* of about 0.8 nm$^{-1}$. This result is remarkable in that the length-dependence of charge transport in a stack of protein films is described by a single exponential attenuation.

To explore what can be the dominating transport mechanism, we also measured the temperature-dependence of the current-voltage characteristics in an experiment for which we had to use evaporated, solid-state Ti/Au top contacts instead of the EGaIn cones, as shown in Figure 1 (b). In contrast to EGaIn, literature reports of evaporated top contacts in protein junctions are scarce and, specifically for bR, limited to Pb/Au and Ti/Au.[47–49] For low temperatures, the conductance is found to be almost independent of temperature until ~50 K, above which the current change can be approximated as exponential, corresponding to an activation energy of roughly 100 meV. We discuss this temperature effect in terms of various theoretical nanoscale charge transport models and examine the role of the interfaces in the long-range charge transport mechanism through the entire bR molecule layer stack.

We recently reported charge transport studies on a covalently stacked multilayer bR system, up to 60 nm in thickness, with decay constants of similar, small magnitude (maximum 0.53 nm$^{-1}$).[48] Importantly, the present findings agree well with those earlier results and complement them, as results are consistent, notwithstanding significant differences in sample architecture and preparation protocols. Sample quality is evaluated with several different, complementary characterization and analysis techniques. We take our results as a clear indication that efficient long-range charge transport persists in electrostatically coupled bR layers up to several tens of nanometers thickness. The transport mechanism can be described mathematically by non-resonant, inelastic tunneling in that it is exponentially length-dependent (though weakly) and only weakly temperature-dependent. Eventually, though, a microscopic explanation for this remarkable finding is still lacking, as coherent tunneling over distances >10 nm appears to be physically unrealistic.

## 2. Results

### 2.1. Surface characterization

*2.1.1. Morphological analysis*

The prepared bR layers were characterized with respect to their roughness and thickness using AFM. **Figure 2** shows tapping-mode images from samples previously functionalized with AhexPA linkers, with zero, one, two, and three layers of bR vesicles (Figure 2 (a-d), respectively). The results of the morphological analysis are summarized in **Table 2**. Bare TiN displays a flat surface with an average root-mean-square (RMS) roughness of $R = 0.4 \pm 0.1$ nm, evaluated from 5×5 µm² images (**Figure S1**). The surface roughness is preserved after linker grafting ($R = 0.4 \pm 0.1$ nm)). The protein layers, as shown in Figure 2 (b), apparently consist of individual bR-lipid vesicles in the form of circular patches. Consequently, the roughness of the single bR layer is increased compared to the substrate ($R = 2.2 \pm 0.3$ nm). Closer inspection via high-resolution AFM and TEM (**Figure S2** (a) and (b), respectively) revealed a patch diameter of roughly 50 nm. The described vesicle appearance and size are in agreement with previous reports for bR.[33,50] TEM images (Figure S2 (b)) provided further insight into the structure of the surface-deposited vesicles, revealing a light (high electron intensity) region in the middle of the vesicle, surrounded by a darker (lower electron intensity) region. As additional layers of bR are grafted (Figure 2 (c-d)), the surface RMS roughness further increases to $R = 3.7 \pm 0.2$ nm for bR bilayers and to $R = 4.1 \pm 0.2$ nm for bR trilayers.

**Table 2**. RMS roughness and mean layer thickness of different organic layers, measured with AFM.

| Sample | RMS roughness[a] [nm] | Mean layer thickness[b] [nm] |
|---|---|---|
| TiN | 0.4 ± 0.1 | - |
| AhexPA linker | 0.4 ± 0.1 | 0.8 ± 1.0 |
| bR single | 2.2 ± 0.3 | 7.5 ± 2.0 |
| bR double | 3.7 ± 0.2 | 12.2 ± 3.5 |
| bR triple | 4.1 ± 0.2 | 15.5 ± 3.8 |

[a] The error of the RMS roughness is the standard deviation over several scans on several samples of the same type; [b] The error of the thickness, $\sigma_d$, is calculated from $\sigma_d = \sqrt{\sigma_1^2 + \sigma_2^2}$, where $\sigma_1$ and $\sigma_2$ are the standard deviations of the two height distributions.

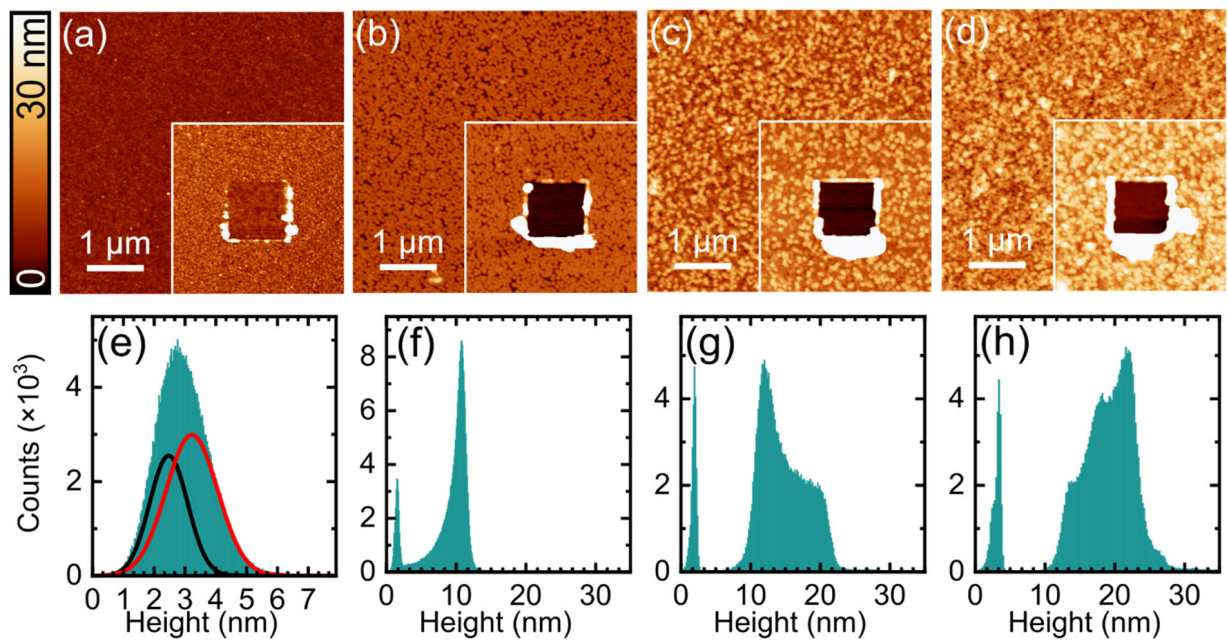

**Figure 2**. (a-d) Typical AFM micrographs of TiN substrates functionalized with AhexPA linkers, (a) before grafting bR and after grafting (b) one, (c) two, and (d) three layers of bR. The insets show AFM scans from the respective sample types where the organic layer has been mechanically removed with the AFM tip in the middle of the scan window (AFM scratching). (e-h) Height histograms of the insets in (a-d), respectively. The black and the red curve in (e) are fitted Gaussian peaks corresponding to the substrate and the linker layer, respectively.

The insets of Figure 2 (a)-(d) show the scratched images of the same sample types (zero to three layers of bR) and Figure 2 (e)-(h) show the corresponding height histograms of the scratched images. The height histograms were used to measure the thickness of the linker and protein layers. In general, the substrate exposed in the scratched area (darker square in the middle of the scans) and the intact organic layer (brighter area around the scratched square) appear as two different distributions in the height histogram. As apparent from Figure 2 (g) and (h), the distributions for the bR bi- and tri-layer samples feature certain substructures ("shoulders", reminiscent of double or triple peaks, respectively), which can be assigned to their particular layer structure: as the second and third layer are less densely packed, the underlying vesicle layer structure can still be identified in the histograms. Notwithstanding that, we define here the mean (total) organic layer thickness $d_{\text{tot}}$ as the difference between the mean heights of the two height distributions. The net protein layer thickness (including the PDDA linker layers) is $d = d_{\text{tot}} - d_{\text{linker}}$. To measure the AhexPA linker layer thickness $d_{\text{linker}}$, since the two distributions were largely overlapping, the histogram was deconvolved with two Gaussian curves (Figure 2 (e)). This analysis yielded the layer thicknesses $d_{\text{linker}} = 0.8 \pm 1.0$ nm for the AhexPA layer, $d = 7.5 \pm 2.0$ nm for the bR vesicle monolayer, $d = 12.2 \pm 3.5$ nm for the bR vesicle bilayer, and $d = 15.5 \pm 3.8$ nm for the bR vesicle trilayer. Note that the protein multilayer thicknesses are not integer multiples of the monolayer thickness, as would be expected for ideal multilayer growth. We attribute this discrepancy to the decreasing packing density (and thus, coverage) of the second and third layers, as seen in the histograms of Figure 2 (f-h). Lower coverage translates to lower mean height values and therefore to lower measured thicknesses for the bi- and tri-layers, even if the individual vesicle dimensions remain the same. When AFM data for a bR layer before and after coating with PDDA were compared (data not shown), no change in roughness or thickness could be resolved, from which we conclude that the polymer linker layer on the bR layer is < 2 nm thick.

The extracted thickness of roughly 7.5 nm of the bR monolayer is significantly higher than the known thickness of the purified purple membrane (~4.5 nm, determined via cryo-electron microscopy[51]), suggesting that vesicle fusion is incomplete and that vesicles collapse into bilayer patches[52] (two membranes upon each other, cf. schematics in Figure 1 (e)), rather than fusing into perfect monolayers, as anticipated earlier.[32] The two differently toned vesicle regions in the negatively stained TEM image of Figure S2 (b) support this conclusion. Therein, the lighter areas of the vesicles may be attributed to protein bilayers and the darker regions to protein monolayers. This would suggest that the vesicles collapse on the solid substrate, break open at their perimeter, and fuse partially. The above analysis was performed on planar samples without a top contact; cross-sectional TEM of a complete TiN/bR/Ti/Au junction (Figure 1 (d)) indicated that the organic layer thickness was reduced slightly to 6-7 nm upon metal evaporation. In neither of the TEM analyses any clear contrast resolving individual bRs was achieved, though, which matches previous studies[50] and may indicate the anticipated high bR density in the partially delipidated protein-lipid vesicles.

### 2.1.2. Surface coverage and elemental analysis

The protein-coated and reference samples were analyzed using Fourier-transform infrared spectroscopy (FTIR) and X-ray photoelectron spectroscopy (XPS) to verify the self-assembly of the molecular layers on the TiN surface and to quantify their composition. In particular, FTIR was used to detect the amide vibrations, which are associated with peptide bonds and are thus a signature of proteins.[53] The FTIR absorption spectra for the amide I and II bands of a TiN/AhexPA sample with and without a single bR layer are shown in **Figure S3**. The spectra display pronounced amide I and II peaks at 1662 cm$^{-1}$ and 1544 cm$^{-1}$, respectively, only after protein grafting, confirming the presence of peptides originating from bR in the grafted film.[54] XPS provides information about the atomic concentration of the elements (proportional to peak area) and the chemical state of the atoms (peak position). The obtained high-resolution photoelectron spectra for the elements C, Ti, and N of one bare O-plasma-treated TiN sample and of TiN samples coated with AhexPA linkers and bR are shown in **Figure 3**. The spectra for all elements (C, Ti, N, O, and P), along with the resulting atomic concentrations, are provided in the Supporting Information (**Figure S4** and **Table S2**). By studying the screening of the Ti 2p signal intensity by the overlying protein vesicle layers (see Figure 3 (b)), we estimated the surface coverage of the bR vesicular layer over the analysis area of XPS (~10 mm$^2$). Here, we define the surface coverage as the ratio of the area covered with protein vesicles (regardless of stack height) to the total sample area. The detailed calculation method is laid out in Section 3 of the Supporting Information. The results are summarized in **Table 3**. We derived surface coverages of roughly 80% for the single bR vesicle layer and > 90% for thicker layers. These results constitute lower boundaries due to the small but non-zero photoelectron transmission through the vesicles, which is neglected in our analysis (see discussion in the Supporting Information).

Furthermore, information gathered with XPS was used to estimate the protein content of the bR vesicles, which also contain a small amount of structural lipids.[55] The C 1s and N 1s spectra (Figure 3 (a) and (c), respectively) exhibit amide (C(=O)-N) features associated with peptide bonds, providing further direct evidence of the presence of protein in the films. More specifically, two peak components are assigned to amides: First, the C 1s peak shoulder appearing only in the protein-coated samples at ~288 eV,[56] left of the main hydrocarbon peak at 284.8 eV (Figure 3 (a) and S4 (a) and (b)). Second, the N 1s peak at ~400 eV,[57] which is more pronounced in the protein samples and varies minimally with the protein layer thickness (see Figure 3 (c)). The deconvolution of the amide peak from the N 1s spectrum enables the estimation of the protein content of the bR vesicles. By considering the atomic composition of the bR molecule (according to PDB code 1R2N) and deducting the contribution of adventitious carbon and the AhexPA linkers from the C, O, N and P atomic concentrations, the protein content of the vesicle (including the PDDA linkers) by atom count and by mass was calculated. The calculation method is explained in detail in Section 3 of the Supporting Information. The results, which are shown in Table 3, suggest a high protein content of > 85% by atom count and by mass in the protein-lipid layer (the rest being lipids, PDDA and structural water). The known protein-lipid ratio in the native purple membrane is 75:25 by mass.[55] The higher protein content observed in this work confirms the partial (>50%) delipidation of the vesicles induced by the detergent OTG. In summary, the structural analysis via XPS revealed the high surface coverage (approximately 80% or higher) and protein content (> 85%) of the bR-lipid vesicle layers. Therefore, we assume that the charge transport discussed hereafter occurs mainly through the bR molecules, as opposed to lipid molecules or conductive defects (metal filaments or other).

**Table 3**. XPS quantitative analysis results from TiN/AhexPA samples with one to three bacteriorhodopsin layers (1-3bR).

| Sample | Surface coverage of bR vesicles[a] [%] | Protein content of vesicles by atom count[b] [%] | Protein content of vesicles by mass[b] [%] |
|---|---|---|---|
| 1bR | 81 ± 9 | 90-100 | 90-100 |
| 2bR | 94 ± 9 | 85-100 | 85-100 |
| 3bR | 91 ± 10 | 90-100 | 90-100 |

[a] The surface coverage is the area covered by bR vesicles divided by the total sample area.; [b] Protein content refers to the bR vesicle layer, including the inter-protein linker layer (PDDA), but excluding the protein-substrate linker layer (AhexPA).

## 2.2 Current-voltage measurements

### 2.2.1. Length-dependence

To investigate the charge transport through the vertical stack of protein layers, current-voltage measurements were carried out at first using EGaIn as top electrode. **Figure 4** (a) displays the mean logarithmic current density $\log|J|$ as a function of the voltage $V$ for bare TiN substrates, for TiN substrates coated with AhexPA linkers, and for those with one to three bR layers, measured at room temperature. Figure 4 (b) displays the same data as in (a), represented on a double-logarithmic scale.

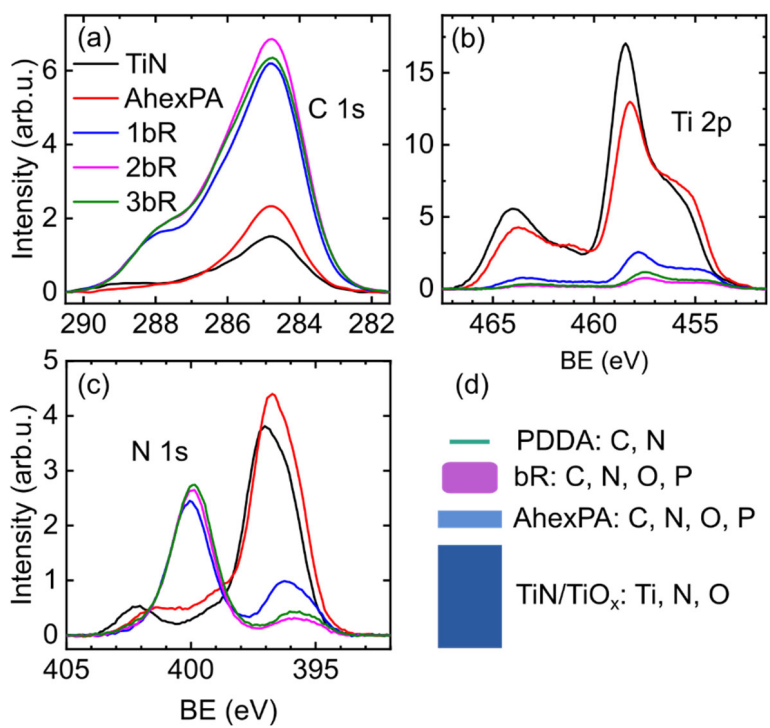

**Figure 3**. (a-c) X-ray photoelectron spectra of (a) C 1s, (b) Ti 2p and (c) N 1s for a blank TiN (black lines), a TiN/AhexPA (red lines), a TiN/AhexPA/1bR (blue lines), a TiN/AhexPA/2bR (magenta lines), and a TiN/AhexPA/3bR sample (green lines). (d) Scheme of the elements (except H) in each material used in this study, following the representation of Figure 1 (c).

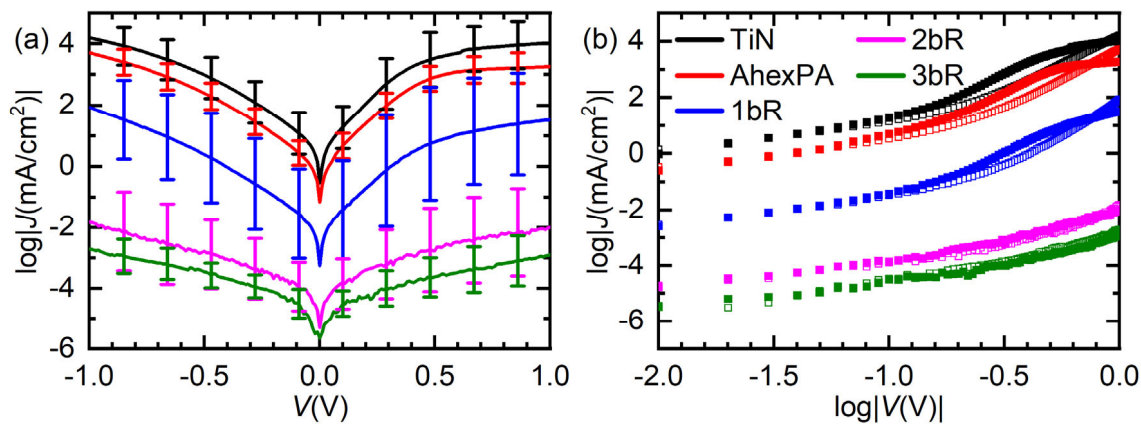

**Figure 4.** Current density-voltage characteristics of TiN (black), TiN/AhexPA (red), TiN/AhexPA/1bR (blue), TiN/AhexPA/2bR (magenta) and TiN/AhexPA/3bR (green) junctions contacted with EGaIn from above at room temperature. (a) Semi-logarithmic, (b) double-logarithmic plot. Shown data are averages from first sweeps, taken on 10-33 different locations on 2-6 sample chips. The error bars in (a) are the standard deviations of the logarithm. In (b), open symbols denote $V < 0$ and closed symbols denote $V > 0$. For raw data, refer to **Figure S6**.

Linear fits to the log|*J*| vs. log|*V*| data between ± 100 mV (corresponding to log|*V*| ≤ −1.0 in Figure 4 (b)) yield slopes close to 1.0 (see **Figure S7** and **Table S4**), meaning that the *J*−*V* characteristics are nearly linear ("ohmic") at low bias. At higher negative bias down to −1.0 V, the characteristics for the junctions with bR turn exponential-like, as can be observed from the cross-over from slope = 1 to slope > 1 in the double logarithmic plot of Figure 4 (b). At higher *positive* bias, the *J*(*V*) characteristics are also exponential-like up to roughly +0.5 V. Above that voltage, for TiN, TiN/AhexPA, and to some extent TiN/AhexPA/1bR, a saturation in the current density is observed, which leads to asymmetric *J*(*V*) traces. As this feature is most pronounced in the reference systems TiN and TiN/AhexPA but vanishes completely after adding two layers of bR, we conclude that it is most likely related to the electrodes and not to the protein layers themselves. Regardless of the shape of the *J*(*V*) curves, we observe that the measured current follows a monotonically decreasing trend with increasing protein film thickness for the entire voltage range under study. More specifically, the bare TiN/EGaIn junction shows the highest current, as expected. The TiN/AhexPA/EGaIn junctions are marginally less conductive. After grafting one layer of bR, the conductance drops by about 2 orders of magnitude, and drops further with every additional bR layer. This trend indicates that the resistance of the proteins (including their interfaces with each other and with the contacts) dominates the total resistance of the protein junctions.

We plotted the natural logarithm of the (geometric contact) area-normalized zero-bias conductance *G* of the different junctions against their thickness *d* to examine how the conductance attenuates with increasing protein layer thickness. Here, all zero-bias conductance values were obtained from fits to the respective *J*(*V*) data in the linear regime, between −50 and +50 mV. The linear trend displayed in **Figure 5** strikingly shows that the conductance of the protein junctions decays practically mono-exponentially (i.e., with a single attenuation constant) with the layer thickness, from the no-protein reference up to the maximum protein film thickness of about 16 nm, within the experimental accuracy. To quantify this behavior, we fitted a linear relationship to the ln*G* vs. *d* data:

$$\ln G = \ln G_\text{C} - \beta d, \tag{2}$$

where $G_\text{C}$ and $\beta$ are defined as in Equation (1). The obtained $\beta$ value at zero bias is $(0.8 \pm 0.2)$ nm$^{-1}$. The same length-dependence analysis was performed for the current density *J* at the voltages *V* = −1.0, −0.9, ..., 1.0 V. The resulting ln|*J*| vs. *d* data also followed a nearly linearly

decaying trend for finite bias (see **Figure S9**), allowing for a voltage-dependent determination of $\beta$. The obtained $\beta(V)$ values all lie between 0.8−1.0 nm$^{-1}$, being lowest between $V = \pm 0.1$ V and slightly increasing to their maximum values at around −0.8 V and +0.5 V, respectively, within the error margin (see **Figure S10**).

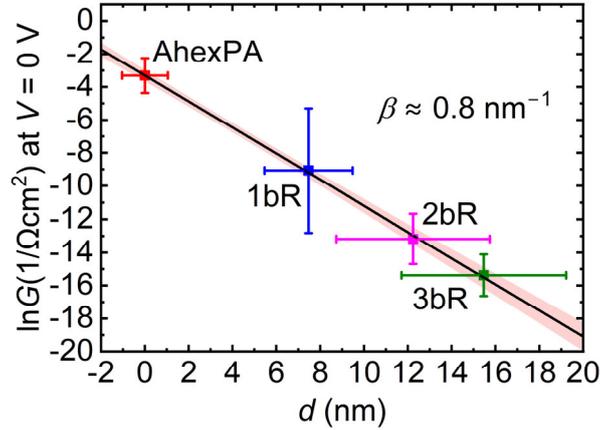

**Figure 5.** Mean natural logarithm of the zero-bias, area-normalized conductance $G$ of TiN/AhexPA (red square), TiN/AhexPA/1bR (blue square), TiN/AhexPA/2bR (magenta square) and TiN/AhexPA/3bR (green square) junctions contacted with EGaIn from above at room temperature, plotted against the protein layer thickness $d$ measured with AFM. The surface of the AhexPA linker layer is set to $d = 0$. The black line represents a linear fit through the experimental data points. The error bars denote the standard deviations. The pink band denotes the 95% confidence interval. ln|$J$| vs. $d$ data at non-zero bias are summarized in the SI (Figure S9).

The obtained value of $\beta$ is lower than the previously reported estimates for bR vesicle monolayers. More specifically, Ron et al.[35,36] reported $\beta = 1.2$ nm$^{-1}$ evaluated at $V = \pm 1$ V for bR monolayers bound on Si/SiO$_x$ via silane linkers and measured with an Hg droplet as top electrode. We later reported similar values for bR monolayers on TiN contacted from above with EGaIn ($\beta = 1.3 \pm 0.4$ nm$^{-1}$) and with evaporated Ti/Au top contact ($\beta = 1.5 \pm 0.4$ nm$^{-1}$), both evaluated at zero bias.[49] It should be noted, however, that those results were derived from two data points only, i.e., from the measured current before and after grafting one layer of bR, which possibly resulted in a larger uncertainty. In the recent work of Bera et al. on bR multilayers,[48] even lower $\beta$ values were reported (0.25-0.5 nm$^{-1}$) for single to triple bR vesicle layers, though with covalent protein-protein coupling and larger measured layer thicknesses. The parameter $G_C$ of Equation (1) is related to the contact resistance. From the linear fit to the

ln$G$ vs. thickness data of Figure 5, we extrapolated a contact (zero length) resistance of 27 Ωcm$^2$ (normalized per contact area) or 177 kΩ (in absolute terms, for our junction area). Ron et al. previously published a contact resistance value of the order of 10$^{-3}$ Ωcm$^2$ (4 orders of magnitude smaller), measured with bR on an amino-terminated linker layer on Si/SiO$_x$, with a Hg top electrode, at ±1.0 V.[36]

Our results point at two major findings: 1) long-range charge transport through electrostatically coupled bR is possible, and it is likely more efficient than originally suggested,[35,36,49] as reflected in the very small $\beta$, and 2) the attenuation of the conductance with distance (= thickness of the protein layer) follows a *mono-exponential* dependence.

*2.2.2. Temperature-dependence*

To further elucidate the charge transport mechanism in electrostatically coupled bR, we fabricated permanent TiN/protein/Ti/Au solid-state devices to carry out temperature-dependent *I-V* measurements inside a vacuum cryo probe station. The measured *J-V* curves for a single-layer bR junction and a bare TiN reference junction at different temperatures are shown in **Figure S8. Figure 6** (a) displays the mean logarithmic current density at low bias ($V = 0.1$ V) for a single-layer bR junction compared to the data for a bare TiN reference junction over a wide range of temperatures, as a function of 1000/*T*. We note that solid-state top-contact bR double and triple layer samples yielded low currents with high noise (data not shown). For the single bR sample, we observed almost temperature-independent transport at low temperatures (below ~50 K) and a moderate increase of ln*J* with increasing temperature above that threshold. To gain insight into the transport mechanism, we modelled the data of Figure 6 with respect to their temperature-dependence in the framework of the four theoretical models introduced in the introduction and in more detail in the Supporting Information: tunneling, thermally activated hopping, variable-range hopping, and transport involving carrier cascade processes.[58] The key equations of each transport model are summarized in Table 1.

- Tunneling: Equation (S4) was adjusted to express the current density, yielding the fitting equation

$$J(T) = J_{0K} + \frac{b_t}{k_B T} \text{sech}^2 \left( \frac{\phi}{2k_B T} \right), \qquad (3)$$

with $J_{0K}$, $b_t$, and $\phi$ as fitting parameters.

- Thermally activated hopping: Equation (S6) was used for the current density, with the addition of a temperature-independent term to account for the constant current at low temperature. This leads to the fitting equation

$$J(T) = J_{0K} + b_h \exp\left(-\frac{\phi}{k_B T}\right), \quad (4)$$

with $J_{0K}$, $b_h$, and $\phi$ as fitting parameters. The zero-temperature current density $J_{0K}$ may originate from a different, temperature-independent mechanism.

- Variable-range hopping: Analogously to conventional hopping, we modified Equation (S10) to obtain the fitting equation

$$J(T) = J_{0K} + b_{vrh} \exp\left[\left(-\frac{\phi}{k_B T}\right)^{\frac{1}{D+1}}\right], \quad (5)$$

with $J_{0K}$, $b_{vrh}$, and $\phi$ as fitting parameters. The exponential parameter $D$ is set to 1, 2, or 3 (for 1D, 2D, or 3D transport, respectively).

- Carrier cascade: Equation (S11) and (S12) were used for the current density, yielding the fitting equations

$$J(T) = J_{0K} + J_{LT} \exp\left(-\frac{\Delta_{LT}}{k_B T}\right), \quad (6)$$

$$J(T) = J_{HT} \exp\left(-\frac{\Delta_{HT}}{k_B T}\right), \quad (7)$$

with $J_{0K}$, $J_{LT}$, $J_{HT}$, $\Delta_{LT}$, and $\Delta_{HT}$ as fitting parameters.

We applied all four different models, expressed by Equation (3)-(7), to fit the ln$J$ vs. 1000/$T$ data in Figure 6 (a). This allowed at first for extracting and comparing the characteristic energy barriers $\phi$. Setting $J_T = J - J_{0K}$, with the zero-temperature current density $J_{0K}$ being directly extracted from the $T$-dependent data of Figure 6 (a), further allowed for fitting the ln$J_T$ vs. 1000/$T$ data as plotted in Figure 6 (b), to extract values for $\Delta_{LT}$ and $\Delta_{HT}$ (cascade model). Here, as a transition in the slopes appears around 150 K (cf., Figure 6 (b)), the 30 K – 130 K interval was selected as the low-$T$ fitting range and the 210 K – 300 K interval was selected as the high-$T$ fitting range. Both characteristic energies were derived for each junction type.

The Mott variable-range hopping model for $D$ = 1, 2, 3 could not describe our data well, yielding unrealistically high energy barriers (see **Table S5** and **Figure S12**), and is therefore not discussed further.

All extracted characteristic energy parameters $\phi$, $\Delta_{LT}$, and $\Delta_{HT}$ are summarized in **Table 4**. Overall, the values of $\phi$, and $\Delta_{HT}$ lie around 100 meV and the values of $\Delta_{LT}$ around 10 meV. Notably, the fitted transport parameters of the protein junctions do not differ much from those of the reference samples (difference < 30 meV), as will be discussed in the next section.

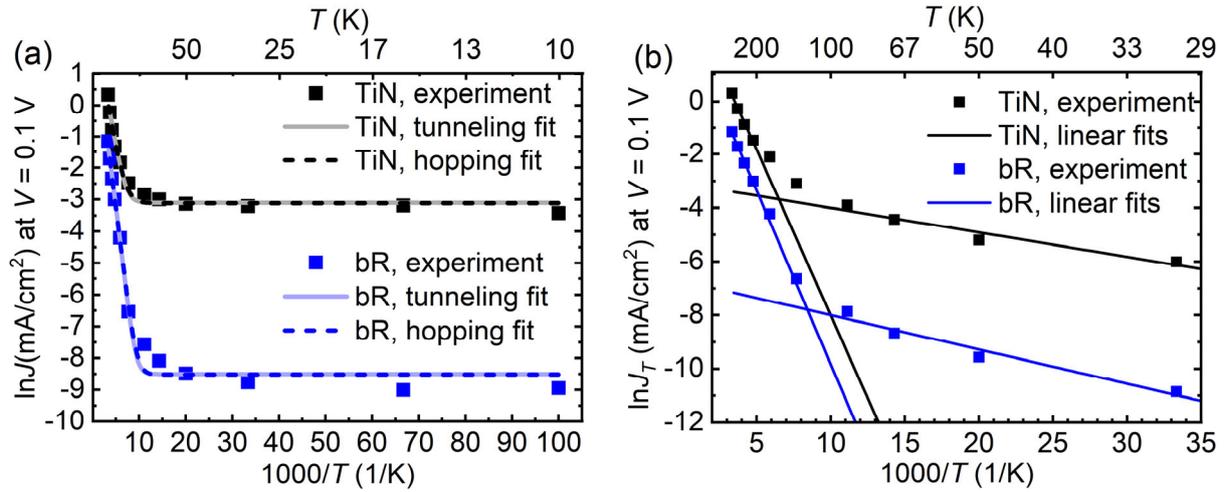

**Figure 6.** (a) $\ln J$ and (b) $\ln J_T = \ln(J - J_{0K})$ at $V = 0.1$ V of the TiN/Ti/Au (black) and of the TiN/AhexPA/bR/Ti/Au (blue) junctions, as a function of $1000/T$. The square data points represent the experimental data. In (a), the solid lines represent fits of the temperature-dependent tunneling model, and the almost coinciding dashed lines represent fits of the thermally activated hopping model to the experimental data. The temperature range is 10 – 300 K. In (b), the lines represent linear fits of the carrier cascade model to the experimental data at low (30 – 130 K) and at high temperatures (210 – 300 K). The temperature range is 30 – 300 K.

**Table 4.** Fitting parameters of hopping, tunneling and cascade transport models to the data of Figure 6.

| Sample | Hopping | | Tunneling | | Carrier cascade | | | |
|---|---|---|---|---|---|---|---|---|
| | $\phi$ (meV) | ± | $\phi$ (meV) | ± | $\Delta_{LT}$ (meV) | ± | $\Delta_{HT}$ (meV) | ± |
| TiN | 75 | 9 | 100 | 10 | 8 | 1 | 107 | 8 |
| 1bR | 100 | 10 | 120 | 10 | 11 | 2 | 112 | 5 |

## 3. Discussion

### 3.1 Assessment of different top contact schemes

This work features two different top contact schemes for charge transport measurements on bR junctions: non-permanent EGaIn cone contacts and permanent evaporated Ti/Au microelectrodes. While EGaIn is used in molecular and also protein electronics as a convenient contact method,[46] non-destructive evaporation of top contacts remains a major challenge. Here, we address the feasibility of evaporated, microscale Ti/Au top contacts for bR junctions by comparing their measured current densities with those measured with EGaIn top contacts (in this work) and with those from previous reports on evaporated Pb/Au top contacts.[47,48] The comparison for ±0.1 V and ±1.0 V bias is summarized in **Table 5**. The logarithmic current densities measured in this work with evaporated top contacts at ±0.1 V deviate from the previously reported values (Pb/Au top contacts) by no more than 0.3 orders of magnitude, which is within the experimental error margin (about 1 order of magnitude in this work). At ±1.0 V bias, the deviation becomes 0.6 orders of magnitude. It is further observed that junctions with EGaIn top contacts yield about 1 order of magnitude lower current densities at low bias, but higher current densities at ±1.0 V bias, as their *J-V* curves are steeper. In summary, the current densities measured in this work with Ti/Au top contacts are in reasonable agreement with, both, previous reports using Pb/Au top contacts and measurements with EGaIn top contacts from this work, particularly at low bias. From the comparison between Ti/Au and Pb/Au top contacts, we deduce that, with the limited data available until now, evaporated Ti/Au is a viable alternative to Pb/Au as top contact material for protein junctions. Nevertheless, it is noted that the junctions presented in this work also employ different bottom electrodes, namely TiN covered with an organophosphonate linker layer, in contrast to the previously reported junctions, which employed $Si/SiO_x$ covered with a silane linker layer.[47,48]

**Table 5**: Comparison of the current densities measured in this work (with EGaIn or evaporated Ti/Au top contacts) with previous reports of bR single layer junctions measured with evaporated top contacts at room temperature, for low (±0.1 V) and higher (±1.0 V) bias.

| Reference | Top contact | Geom. area [µm²] | log|*J*[A/cm²]| at -0.1 V | +0.1 V | -1.0 V | +1.0 V |
|---|---|---|---|---|---|---|
| [47] | Pb/Au | 5000 | -3.8 | -3.4 | - | - |
| [48] | Pb/Au | 5000 | -3.7 | -3.7 | -2.6 | -2.6 |
| this work | Ti/Au | 2500 ± 100 | -3.5 | -3.5 | -2.0 | -2.0 |
| this work | EGaIn | 7800 ± 1600 | -4.5 | -4.4 | -1.1 | -1.5 |

### 3.2 Length-dependence of charge transport

In the Results section, four transport models have been implemented to fit our electrical data with respect to their length-dependence and temperature-dependence. The outcome of these fits is summarized in **Table 6** and its implications are discussed hereafter. The observed mono-exponential attenuation of the electrical current as a function of length up to 15.5 nm matches the prediction for inelastic (coherent) tunneling transport (as in Equation (1)), remarkably even with a single (small) decay constant $\beta$. To the best of our knowledge, such a finding with a single $\beta$ has not yet been reported for long-range charge transport in protein layers. The above-mentioned tunneling transport though, is an unrealistic mechanism due to the expected carrier scattering in this statically and dynamically disordered medium as soon as the charge transport length exceeds a certain critical value of typically just a few nm. Even for the very special *multi-heme* cytochrome system this critical distance has been calculated, via density-functional theory, to be no more than 5-7 nm.[59]

Our electrical data on length-dependence are inconsistent with thermally activated and variable-range hopping, which predict a $1/d$ length-dependence for the conductance (see Equation (S5) and (S7)).[14,27] To verify this, we plotted the $G(0\ V)$ vs. $d$ data at room temperature in a double-logarithmic representation (see **Figure S11** and associated discussion in the Supporting Information), which clearly ruled out a $1/d$ dependence of the electrical data. In previous reports on conjugated organic molecules, similar low-$\beta$ tunneling-like transport characteristics have often been interpreted in terms of long-range single-step resonant transport[1,9,12,60] or sequential (multi-step) tunneling.[5,6,10] Often, a characteristic transition from a higher $\beta$ regime (usually assigned to coherent off-resonant single-step tunneling) to a regime with lower $\beta$ (sequential tunneling) was reported. Such a transition is not observed in our study. This suggests that, in our case, the *mechanism is remarkably uniform* across the entire thickness range under study, covering up to 15.5nm, as reflected in a single $\beta$ of ~0.8 nm$^{-1}$ at zero bias. This value is small compared to non-resonant tunneling through thin films of saturated[39,40] and conjugated[19] organic molecules, and even with respect to some of the previously published values for proteins.[35,36,49] To put this in relation using the Simmons model,[61] a $\beta$ of 0.8 nm$^{-1}$ translates into a shallow energy barrier of only about 5 meV, according to Equation (S2), if zero-bias tunneling through a rectangular barrier is considered and if $m^* = m_0$ (electron rest mass) is assumed. If we assume a small effective mass of only 0.04 $m_0$, as Casuso et al. have reported for conductive-probe AFM measurements on bR (purple membrane) monolayers,[62] then the calculated energy barrier becomes approximately 125 meV, which is comparable to the larger barrier values summarized in Table 4. Even with the exact formula of Equation (3) for the tunneling conductance, one obtains an energy barrier $\phi$ only 24% higher than with the approximation of Equation (1), i.e. the barrier height remains in the same order of magnitude.

## 3.3 Temperature-dependence of charge transport

Having excluded thermally activated and variable-range hopping transport due to the measured length-dependence of the conductance, tunneling (Equation (1) and (S4)) is the only mathematical model, out of the four considered, that can provide an accurate fit to both the length and the temperature-dependence of our experimental data. The obtained energy barrier height is about 100 meV (Table 4), i.e., only a few times larger than the thermal energy at room temperature. In an idealized molecular junction, the relevant barrier should correspond to the energy difference between the Fermi level and the closest available molecular level, which can be either the HOMO or the LUMO.[14] Reported HOMO-LUMO gaps in bR, with levels in part corresponding to the retinal, are in the range of 2-3 eV.[48,63] This however does not necessarily contradict our estimate of at least one order of magnitude lower barriers, if the HOMO or the LUMO level lies close to $E_F$.

It should be noted that the obtained energy barrier height does not differ significantly between the sample with protein layer, and the bare reference junction (see Table 4). Consequently, the thermal activation of the charge transport may, in fact, be dominated by the electrodes or their interfaces to the electrostatically bound proteins, *in which case the transport through the actual protein layer appears to be nearly activation-less*. This reasoning is in line with the recent findings of Bera et al. for covalently linked bR multilayer junctions,[48] which in turn also implies that the observed electrode predominance additionally appears to be insensitive to the actual bottom and top electrode material combinations and the type of electrode-molecule coupling (covalent or electrostatic) – as those of Bera et al. differed from our work.

**Table 6**: Fitting outcome of different charge transport mechanisms, separated into their length-dependent and temperature-dependent part.

| Mechanism | Fits length-dependence? | Fits temperature-dependence? |
| --- | --- | --- |
| Tunneling | yes | yes, with $\phi \sim 100$ meV |
| Hopping | no, instead $G \propto d^{-9}$ | yes, with $\phi \sim 100$ meV |
| Variable-range hopping (Mott) | no, instead $G \propto d^{-9}$ | yes, with high $\phi$ values ($\sim 2 - 5000$ eV) |
| Carrier cascade | - | yes, with $\Delta_{LT} \sim 10$ meV and $\Delta_{HT} \sim 100$ meV |

The energy values obtained from the carrier cascade model may provide additional information about the relative position $\Delta_{LT}$ of the next nearest occupied or unoccupied levels with respect to HOMO and LUMO, respectively. The obtained values $\Delta_{LT} \approx 10$ meV suggest a minimal spacing of energy levels outside the HOMO-LUMO gap, surprisingly also (again) for the reference junction without protein, which requires further investigation.

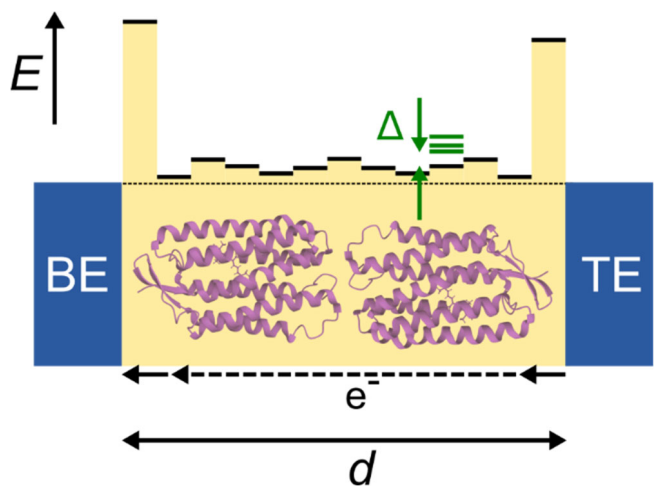

**Figure 7.** Schematic illustration of a multi-barrier tunneling charge transport mechanism through a bR vesicle, comprising two bR molecules back-to-back. The dark blue blocks represent the electrodes (bottom electrode (BE): TiN, top electrode (TE): EGaIn). Linkers between the BE and the protein, as well as lipids, are not shown for clarity. The black lines represent a possible qualitative distribution of near-resonance energy levels of the entire layer stack, here exemplarily the LUMO levels. For one of them, a few neighboring levels above the LUMO are included in green color, as an example. Narrow, high barriers are present at the interfaces to the contacts. The dotted line is the Fermi level. The (bottom) black arrows indicate the possible electron transport sequence. Solid arrows indicate the tunneling steps through the protein-linker-contact interfaces. The dashed arrow indicates the unknown transport mechanism through the proteins.

**3.4 Transport mechanism considerations**

A plausible, qualitative charge transport model for the studied single-layer bR system summarizes the abovementioned findings and is illustrated in **Figure 7**. Based on the temperature-dependent analysis results, we hypothesize that the energy landscape contains high and narrow injection tunneling barriers of ~100 meV at the contacts, which likely limit the measured conductance of the system and determine its (mild) temperature-dependence. Then, the protein layer contributes a very low resistance to the overall junction, which – together with the electrostatic protein-protein interfaces – could explain the low determined $\beta$ value. The difference in the conductance of the protein-less reference junctions and the single-layer bR junctions may be explained by the different number of interfaces present therein (one vs. two, respectively). The interpretation of charge transport through the inner part of the protein in terms of established mechanisms, however, remains elusive. While alternative hypotheses for transport cannot be definitively ruled out, the interface-limited one is compatible with the observed temperature and length-dependence, it is physically realistic (in contrast to single-step coherent tunneling through the entire junction) and in agreement with the previous conclusions of Bera et al.[48]

We discuss below the implications of the through-interface-tunneling hypothesis for the estimated value of $\beta$. The extracted current attenuation constant $\beta$ is an effective (mean) value over the entire thickness of the investigated layer stack. If we consider, for example, just two individual rectangular tunneling barriers of equal length $L$ and equal electron effective mass inside, but of (in general) different heights $\phi_1$ and $\phi_2$, in series, the conductance for tunneling through this double barrier will still obey Equation (1) derived from the Simmons model, with

$$\beta = \frac{\beta_1 + \beta_2}{2}, \tag{8}$$

now being the mean attenuation constant. Here, $\beta_1$ and $\beta_2$ are attenuation constants associated with each individual barrier. The effective barrier height for the double barrier is given by

$$\phi = \left(\frac{\sqrt{\phi_1} + \sqrt{\phi_2}}{2}\right)^2 \tag{9}$$

now representing the effective barrier height of the entire junction, according to Equation (S2). From Equation (1) and (S2), it follows that if the total tunnel barrier length $2L$ is overestimated by a given factor, then the decay coefficient $\beta$ is underestimated by the same factor and the barrier height is underestimated by the square of that factor. In our extraction of $\beta$ (Section 2.2.1), we identified the tunneling barrier length, here named $2L$, with the respective, total geometrical junction length $d$. A presumption of short, but high tunneling barriers of length $L$ directly at the interfaces to both contacts (each characterized by a $\beta$ value much larger than 0.8 nm$^{-1}$), combined with the inner protein segments contributing with only negligible resistivity ($\beta$ negligibly small), however, may result in the same overall effective, measured value for $\beta$ (0.8 nm$^{-1}$). In other words, we assign high and short tunnel barriers at the contacts to dominate the overall resistance with a large $\beta$ each, values, which are underestimated by the effective $\beta$ value as determined for the entire junction.

Finally, what causes the current to decay mono-exponentially beyond the single-layer thickness, as observed from the length-dependent measurements with EGaIn top contacts? Since temperature-dependent measurements suggest that transport is interface-limited, we hypothesize that the observed mono-exponential decay originates from the electrostatic protein-protein interfaces, which include the cationic PDDA linker layers. These interfaces are assumed to add barriers to the transport process as the junction length increases (not shown in Figure 7). The additional protein-protein barriers may co-determine the effective barrier height of the entire junction along with the protein-electrode barriers discussed above.

## 4. Conclusions

In summary, we have studied junctions of bR stacked multilayers sandwiched between a TiN bottom electrode and an EGaIn or a Ti/Au top electrode. We found high protein content (>85%) of the protein-lipid vesicles via XPS analysis, which supports the common assumption that the measured current flows predominantly through the stacked bR molecules and not through the lipids. Overall, long-range charge transport through electrostatically coupled, multilayered bR appears to be very efficient, with currents decaying mono-exponentially as a function of distance over approx. 16 nm thick junctions, with a small current decay constant $\beta \sim 0.8$ nm$^{-1}$ at zero bias. Mild dependence of the current on temperature was observed above 50 K. Below 50 K, the conductance appears to be temperature-independent. Both, the measured length and temperature-dependence of the current match the ones predicted for non-resonant, inelastic (coherent) tunneling, a mechanism which, however, is unphysical over the investigated distances, from today's generally accepted conviction. Therefore, the microscopic mechanism behind the apparently very efficient transport mechanism within the proteins remains under further investigation and calls for new theoretical approaches.

Notwithstanding that, our findings indicate that tunneling through interfacial barriers dominate much of the characteristics. The mild temperature-dependence of the current on temperature is likely limited by tunneling through such energy barriers at the protein-electrode interface(s), whereas the transport mechanism through the inner protein structure itself appears to be more efficient than that across the contact interfaces. The mono-exponential current dependence on length with a low attenuation coefficient throughout the entire length range (up to 15.5 nm) is possibly explained by transport barriers at the electrostatic protein-protein interfaces. Future experimental work shall focus on developing complementary means to elucidate the processes that enable such efficient transport through the inner protein molecules. These means may include, e.g., a three-terminal protein nanodevice to facilitate electrostatic gating, thereby gathering crucial information about the protein energy level distribution.

## 5. Experimental Section

*Sample preparation*: Figure 1 (c) provides a schematic side view of the sample structure and the electrical device configuration. Silicon wafers (diameter 200 mm, orientation (100), p-type doping with boron, resistivity 1−12 Ωcm) were thermally oxidized to grow a ~850 nm thick $SiO_2$ layer and subsequently sputter-coated with a 50 nm thick titanium nitride (TiN) layer. Two types of substrate chips, diced from the full wafers, were prepared: a) Flat, planar chips (10 × 10 $mm^2$) with homogeneous TiN layer, and b) patterned chips (8 × 8 $mm^2$), where the TiN layer was structured into bottom electrode microstructures using stepper i-line photolithography and dry plasma etching. AlSi contact pads were additionally deposited to facilitate contact with electrical probe needles. For more details of bottom contact geometry and fabrication refer to Reference [64].

For cleaning and surface activation, all chips were ultrasonicated successively in acetone, isopropanol, and water and subjected to short oxygen plasma treatment for 30 s at 80 W power and 0.3 mbar oxygen pressure. A layer of 6-aminohexylphosphonic acid (AhexPA; Figure 1 (c); purchased in chloride form from Sikemia, Grabel, France) was grown via self-assembly from 1 mM solution in isopropanol (3 h immersion) and annealing on a hotplate at 130° C for 1 h under ambient atmosphere.[49] The phosphonic acid self-assembled monolayer (SAM) served as a linker for electrostatically grafting bR molecules on the titanium nitride thin native oxide surface. bR (Figure 1 (e)) was extracted from native membrane fragments of *Halobacterium salinarum* via detergent treatment in octylthioglucoside (OTG) to form protein-lipid vesicles, which are partially delipidated compared to the native purple membrane. The vesicles were then dispersed at a concentration of 4 µM in phosphate-buffered saline solution, according to a previously published protocol.[30] The OTG-treated bR dispersion was used to prepare self-assembled protein vesicle layers on the functionalized TiN surface via incubation for 30 min and subsequent immersion in deionized water for at least 3 h. The above process exploits the net negative surface charge of bR under nearly pH neutral experimental conditions (isoelectric point of bR: ~5),[65] to immobilize a single thin bR vesicle layer on the (positively charged) amino-terminated TiN surface (see AFM image in Figure S2 (a)). Stacked layers of bR were prepared by using the cationic polymer poly-diallyl-dimethylammonium (PDDA; Figure 1 (e); purchased in chloride form from Sigma Aldrich (St. Louis, United States) as an electrostatic vesicle-vesicle linker.[24] Specifically, the previously protein-coated chips were immersed in 0.2 %v/v PDDA and 0.5 M NaCl aqueous solution for 15 min, rinsed with deionized water, and dried under a $N_2$ stream. To grow an additional bR layer, the above-described bR grafting protocol was repeated. Double and triple layers of bR were prepared in this study. Bare plasma-treated TiN and AhexPA-coated TiN samples were used as references. On unpatterned substrates, conical EGaIn tips were used as a direct top contact (see Figure 1 (a)). On pre-patterned substrates comprising structured bottom electrodes, after protein grafting, a nickel shadow mask was aligned before the deposition of 10 nm Ti and 80 nm Au via electron-beam evaporation inside an L560 system (Leybold, Cologne, Germany), at a pressure of ~$10^{-6}$ mbar. Evaporation rates were kept low (Ti: 0.5 – 1 Å/s, Au: 1- 2 Å/s) to minimize damage to the protein layers. This fabrication process yielded permanent TiN/AhexPA/bR/Ti/Au crossbar devices with junction area (400 ± 40) µm$^2$ or (2500 ± 100) µm$^2$.[64] A cross-sectional TEM image of such a solid-state protein junction is shown in Figure 1 (d).

For TEM analysis of surface-deposited vesicles (without top contact), bR vesicles were grafted onto formvar/carbon-coated copper TEM grids (Electron Microscopy Sciences, Hatfield, Pennsylvania, United States) according to the following adjusted bR grafting protocol: The grids were first subjected to mild oxygen plasma treatment for 45 s at 0.7 mbar oxygen pressure. Then, their surface was activated in 10 mM $N$-ethyl-$N'$-(3-(dimethylamino)propyl)carbodiimide and $N$-hydroxysuccinimide (EDC and NHS) aqueous solution for 20 min, rinsed with fresh DI water, and incubated with bR vesicle dispersion (see above) for 30 min. Finally, the grids were covered with a drop of DI water and incubated for 1.5 h, before being rinsed with fresh DI water and dried at the end. This protocol yielded sparse layers of bR vesicles on the carbon surface of the grids. To optimize the TEM image quality, the specimens were stained with uranyl formate aqueous solution right before inspection.

*Atomic force microscopy*: Atomic force microscopy (AFM) was used to assess the morphological characteristics of the bR layers using a Dimension V AFM (Bruker/Veeco, Billerica, Massachusetts, United States) operated in tapping mode. Two different probe types were used: Tap190DLC diamond-like-carbon-coated tips (force constant 48 N/m, resonance frequency 190 kHz, tip radius 15 nm) from BudgetSensors (Sofia, Bulgaria) and Super Sharp

Silicon tips (same force constant and resonance frequency, tip radius < 2 nm) from Nanosensors (Neuchatel, Switzerland). In order to measure the bR layer thickness, the so-called AFM "scratching" technique was employed, in which the AFM is operated in contact mode to mechanically remove the organic layers from a small area of the substrate.[66] A subsequent tapping-mode scan of the scratched area allows for the layer thickness determination by measuring the height difference between scratched and unscratched areas. The spring force exerted from the cantilever to the substrate was tuned in such a way that the tip fully removed the organic material, but did not damage the harder underlying substrate. Image acquisition was carried out with Nanoscope 7 software. Image processing and analysis were carried out with Nanoscope, WSXM,[67] and Origin software.

*Electrical measurements*: Electrical measurements were performed using EGaIn top contact junctions and permanent solid-state junctions. On planar, unpatterned samples, current-voltage (IV) measurements were performed at ambient conditions on dry samples with a custom-built setup that uses EGaIn to apply a soft top contact to the organic layers.[45,46] This formed a vertical protein junction between the TiN bottom electrode and the EGaIn top electrode (Figure 1 (a) and (c)). The diameter of the EGaIn contact was $(100 \pm 10)$ $\mu$m, which translates to a geometric contact area of $(7.8 \pm 1.6) \cdot 10^3$ $\mu m^2$. Data were typically acquired in a cyclic manner, i.e., with the applied bias voltage ramping up from 0.0 V to a minimum of −1.0 V, from there up to the maximum bias +1.0 V, and finally back to 0.0 V, before starting a subsequent cycle. Bias was applied in a floating configuration, with the positive pole connected to the bottom contact, as shown in Figure 1 (a). The voltage sweep rate was 20−40 mV/s unless otherwise noted. We note that the measured current magnitudes typically varied monotonically from cycle to cycle (see Figure S5). More specifically, current magnitudes increased from cycle to cycle at maximum positive bias and decreased from cycle to cycle at maximum negative bias. The zero-bias resistance depended both on the sweep direction (higher resistance at negative sweep rate) and on the cycle number (nearly exponential increase). Such variability of the electrical conductance during cycling in junctions with conical EGaIn tips has been reported before and is commonly attributed to the metastable nature of the EGaIn tips.[68] As this effect occurs for the reference (linker only) samples as well as for the linker/bR samples, we can further exclude that it was related to the protein layer conductance possibly changing with time/cycle. To ensure reproducibility and acquisition of fully comparable data, only the measurement traces from each first full sweep are always shown and used for analysis and discussion. Additionally, *I-V* traces with apparent short-circuits ($|I| > 10$ mA) or current leaps higher than 2 orders of magnitude (dlog$|I| > 2$) in their first sweep, were excluded from the analysis. Permanent top-contact protein junctions were measured at variable temperatures inside a CRX-6.5K vacuum cryo-probe station (Lake Shore Cryotronics, Westerville, United States), cooled with a closed-cycle He refrigerator, equipped with tungsten probes, at $\sim 10^{-6}$ mbar pressure, down to nominally 10 K. The geometric contact area of these junctions was $50 \times 50$ μm$^2$ = 2500 μm$^2$. Voltage cycling was performed in the same fashion as with EGaIn top contacts. In either measurement scheme, the device was connected to a Keithley 2635B source-measure unit (Tektronix, Beaverton, Oregon, United States) with triaxial cables, to measure the DC current through the organic layers. Data acquisition was carried out with custom-written LabVIEW or Python measurement scripts. Data analysis was carried out using Python scripts and Origin software.

*X-ray photoelectron spectroscopy*: X-ray photoelectron spectroscopy (XPS) served as an elemental analysis technique, in order to study the composition of the protein layers. Measurements were carried out in a custom-built setup (parts from SPECS Surface Nano Analysis, Berlin, Germany) under ultra-high vacuum ($\sim 6 \cdot 10^{-9}$ mbar). The setup was equipped with an XR 50 X-ray source, operated at a voltage of 12 kV and an emission current of 16.8 mA, and a Phoibos 100 hemispherical electron analyzer with the pass energy set to 25 eV. In this study, a Mg anode was utilized (X-ray photon energy $E = 1253.6$ eV). Photoelectron spectra were acquired at 0.1 eV energy resolution and 1 s dwell time and were averaged over two to six scans. The spectra were analyzed with CasaXPS (charge referencing, background subtraction, fitting, and integration) and Origin software (intensity normalization and plotting). The spectra were charge-referenced with respect to the C 1s hydrocarbon peak at 284.8 eV, background-subtracted, and corrected for analyzer transmission, photoelectron escape depth and transition cross-section. For the latter, relative sensitivity factors (Scoffield cross-sections[69]) from the CasaXPS library were used. Fitting was performed using mixed Gaussian-Lorentzian lineshapes. Plotted data were additionally normalized by the total corrected peak area of the respective sample to ensure comparability between different samples.

*Transmission electron microscopy*: Transmission electron microscopy (TEM) was used to access the structure of the bR vesicles. Top-view imaging was carried out with a 120 kV Tecnai

TEM (FEI, Hillsboro, Oregon, United States). Cross-sectional imaging was carried out by SGS Institut Fresenius (Dresden, Germany).

*Fourier-transform infrared spectroscopy*: Fourier-transform infrared (FTIR) spectroscopy was performed in attenuated total reflection (ATR) mode on TiN samples coated with AhexPA linker layers, with and without bR layers. The method verified the presence of bacteriorhodopsin on the sample and detected possible changes in the protein's secondary structure. Spectra were acquired at room temperature and under 0.2 mbar vacuum with a Vertex 70v spectrometer (Bruker). The spectrometer was equipped with a Globar light source, a KBr beam splitter, and a nitrogen-cooled mercury cadmium telluride (MCT) detector. For the ATR measurements, the sample was pressed upside down against a Ge hemispherical ATR crystal module with a built-in pressure applicator. The absorbance $A$ of the sample was calculated according to the formula $A = \log(I_{ref} / I_{sample})$, where $I_{ref}$ and $I_{sample}$ are the reflected IR signal intensities of the reference and the sample under study, respectively.

## Supporting Information

Supporting Information is available.

## Acknowledgements

We thank the Department of Silicon Technologies and Devices of the Fraunhofer Institute for Electronic Microsystems and Solid State Technologies (EMFT) for providing the TiN substrates, R. Mittermeier for technical support, and S. Saxena, A. Vilan, C. Nijhuis, and J. Blumberger for fruitful discussions. We further acknowledge access to the XPS and FTIR setups of the Walter Schottky Institute, Garching, Germany (Profs. I. Sharp and M. Stutzmann). This work was funded by the German Research Foundation (DFG; grant No. TO266/10-1) and the Federal Ministry of Education and Research (BMBF; grant No. 16FMD01K, 16FMD02, and 16FMD03). Further support at the Weizmann Institute of Science came from the Tom and Mary Beck Center for Advanced and Intelligent Materials. J.A.F. acknowledges the Science and Engineering Board (SERB) of the Department of Science and Technology (CRG/2022/000584), Government of India, for financial support.

Supporting information

for

**Mono-exponential Current Attenuation with Distance across 16 nm Thick Bacteriorhodopsin Multilayers**

*Domenikos Chryssikos, Jerry A. Fereiro, Jonathan Rojas, Sudipta Bera, Defne Tüzün, Evanthia Kounoupioti, Rui N. Pereira, Christian Pfeiffer, Ali Khoshouei, Hendrik Dietz, Mordechai Sheves, David Cahen, and Marc Tornow\*.*

\* Corresponding author: <tornow@tum.de>

**Introduction**

1. This document contains additional theory, experimental data, analyses, figures, and discussion that are supplementary information to the main text. The following information is presented:

    1. **Charge transport mechanism overview**
    2. **AFM and TEM imaging**
    3. **Detailed FTIR and XPS spectra**
    4. **Current-voltage measurements**
    5. **Voltage dependence of current attenuation coefficient**
    6. **Double-logarithmic conductance-length plot**
    7. **Variable-range hopping model fitting**

# 1. Charge transport mechanism overview

In this section, we will review the theoretical length and temperature-dependent characteristics of four fundamental charge transport models used for data analysis in this work, at low bias: temperature-dependent tunneling, thermally-activated hopping, variable-range hopping (VRH), and carrier-cascade transport. The length and temperature dependence of the conductance in each model are summarized in Table 1 of the main text.

*Temperature-dependent tunneling*: In the off-resonant case, coherent quantum-mechanical tunneling transport is characterized by an exponential decay of the current (or of the differential conductance) with distance. As derived in the original Simmons model,[1] in tunneling through a rectangular barrier, in the limit of low applied bias, low temperature, and small transmission, the conductance decays mono-exponentially, according to Equation (1) of the main text, i.e.,

$$G = G_C \exp(-\beta d), \qquad (S1)$$

where $G_C = G(d = 0)$ is the contact (zero-length) conductance[2,3], $d$ is the barrier width, i.e., the layer thickness, and $\beta$ is the current decay coefficient, which is given by the expression

$$\beta = \frac{2\sqrt{2m^*\phi}}{\hbar}, \qquad (S2)$$

where $m^*$ is the charge carrier effective mass, $\phi$ is the energy barrier height and $\hbar$ is the reduced Planck constant. The exact tunneling conductance formula, which also holds at higher transmission (low or short energy barrier), reads

$$G(d) = G_C \left[1 + a \sinh^2\left(\frac{\beta d}{2}\right)\right]^{-1}, \qquad (S3)$$

where $a$ is a dimensionless parameter that depends on the charge carrier energy and the energy barrier height. Modifications of the above equations account for barriers of different shape (e.g., asymmetric trapezoidal[4]), for larger applied bias, or for higher temperature. When the electron thermal energy $k_B T$ becomes comparable to the barrier height, the thermal spread of the Fermi-Dirac distribution $f(E) = \left[1 + \exp\left(\frac{E - E_F}{k_B T}\right)\right]^{-1}$ (where $E_F$ is the Fermi energy), which is typically neglected in Simmons theory, gives rise to the conductance formula [3,5,6]

$$G(T) = G_{0\text{K}} + \frac{B_\text{t}}{k_\text{B}T}\operatorname{sech}^2\left(\frac{\phi}{2k_\text{B}T}\right), \tag{S4}$$

where $G_{0\text{K}}$ is the zero-temperature conductance, $B_\text{t}$ is a parameter with dimensions of conductance × energy that depends on the electrode-molecule coupling, $k_\text{B}$ is the Boltzmann constant, and sech(x) = 1/cosh(x).

*Thermally activated hopping*: Incoherent hopping transport in molecular systems (those lacking electronic band structure) usually involves step-wise sequential tunneling between localized sites within the molecule, where level resonance for tunneling is established by thermal activation.[3] The hopping conductance is inversely proportional to length [7], i.e.,

$$G(d) = \frac{A_\text{h}}{d}, \tag{S5}$$

where $A_\text{h}$ is a parameter with dimensions of conductance × length, and exhibits an Arrhenius-type dependence on temperature according to the formula[3,6,8]

$$G(T) = B_\text{h}\exp\left(-\frac{\phi}{k_\text{B}T}\right), \tag{S6}$$

where $B_\text{h}$ is an effective, infinite-temperature conductance limit. In hopping conduction, the energy barrier is synonymous to the activation energy, often denoted by $E_\text{A}$ or $E_\text{a}$.[6,9,10]

While in practice it is often difficult to assign an experimentally obtained length dependence unambiguously to one of the different transport mechanisms, tunneling and hopping can usually be successfully separated, as demonstrated by AFM measurements on molecular junctions by the Frisbie group.[9] This is owing to their disparate scaling with distance, which for activated hopping is characterized by the inverse proportionality of the conductance to the length of the junction (Equation (S5)) – in stark contrast to the exponential dependence for coherent tunneling (cf. Equation (S1)). Consequently, tunneling is typically the dominant conduction mechanism at short junction lengths and low temperatures, whereas hopping dominates at longer lengths and higher temperatures.[9,11]

*Variable-range hopping:* When (at lower temperatures) the activated process of tunneling to neighboring sites gets less probable, tunneling over longer distances can occur. The theory of VRH was developed by Shklovskii and Efros for lightly doped semiconductors.[12] The transversal VRH conductance is inversely proportional to length, as in thermally activated hopping:

$$G(d) = \frac{A_{\text{vrh}}}{d}, \quad (S7)$$

where $A_{\text{vrh}}$ is a parameter with dimensions of conductance × length. The temperature-dependent expression for the VRH conductance is [12,13]

$$G(T) = B_{\text{vrh}} \exp\left[-\left(\frac{\phi}{k_B T}\right)^p\right], \quad (S8)$$

where $B_{\text{vrh}}$ is an effective, infinite-temperature conductance. The exponent $p$ lies in the range (0, 1) and is given by

$$p = \frac{\mu+1}{\mu+D+1}, \quad (S9)$$

where $\mu$ is the exponent in the power-law dependence of the density of states on energy ($g(E) \propto E^\mu$) and $D$ = 1, 2, 3 is the number of hopping dimensions. In the simple case of energy-independent density of states ($\mu$ = 0), Equation (S8) reduces to Mott variable-range hopping[12,14], where

$$G(T) = B_{\text{vrh}} \exp\left[-\left(\frac{\phi}{k_B T}\right)^{\frac{1}{D+1}}\right]. \quad (S10)$$

*Carrier cascade*: Papp et al. have recently developed a model based on the Landauer-Büttiker formalism,[15] in which charge carriers are assumed to enter the molecule at a higher energy than the LUMO and then transition to lower energies within the molecule in a cascade-like motion.[16] This "carrier cascade" model predicts an Arrhenius-like *T*-dependence of *J* with different parameters in the low and the high *T* regime, respectively. For low *T*, the conductance is given by[15,16]

$$G(T) = G_{0K} + G_{LT} \exp\left(-\frac{\Delta_{LT}}{k_B T}\right), \quad (S11)$$

where $G_{0K}$ is the zero-temperature conductance, $G_{LT}$ is a pre-exponential factor with dimensions of conductance, and $\Delta_{LT}$ corresponds to the energy difference either between the highest occupied molecular orbital (HOMO) and the next lower level ($E_{HOMO} - E_{HOMO-1}$) or the lowest unoccupied molecular orbital (LUMO) and the next higher level ($E_{LUMO+1} - E_{LUMO}$). For high $T$, where $G(T) \gg G_0$, the model yields

$$G(T) = G_{HT} \exp\left(-\frac{\Delta_{HT}}{k_B T}\right), \tag{S12}$$

where $G_{HT}$ is an effective infinite-temperature conductance, and $\Delta_{HT}$ corresponds to the energy difference between HOMO and LUMO ($\Delta_{HT} = E_{LUMO} - E_{HOMO}$).

## 2. AFM and TEM imaging

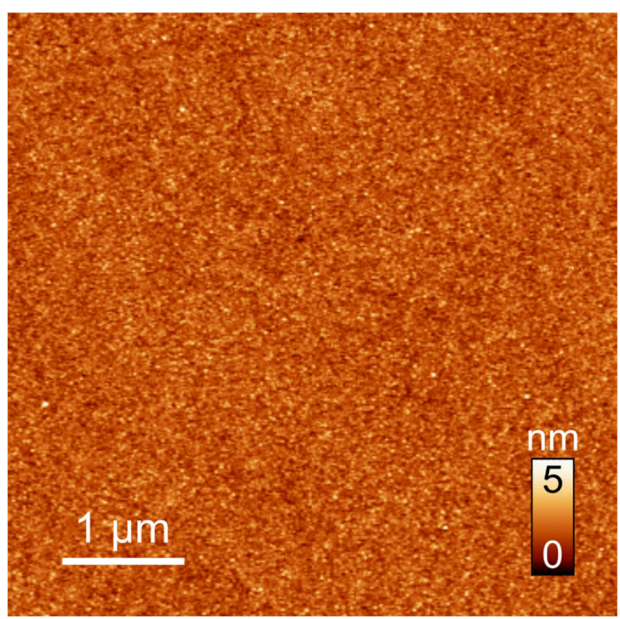

**Figure S1.** Representative AFM image of a solvent-cleaned and O-plasma-treated TiN surface.

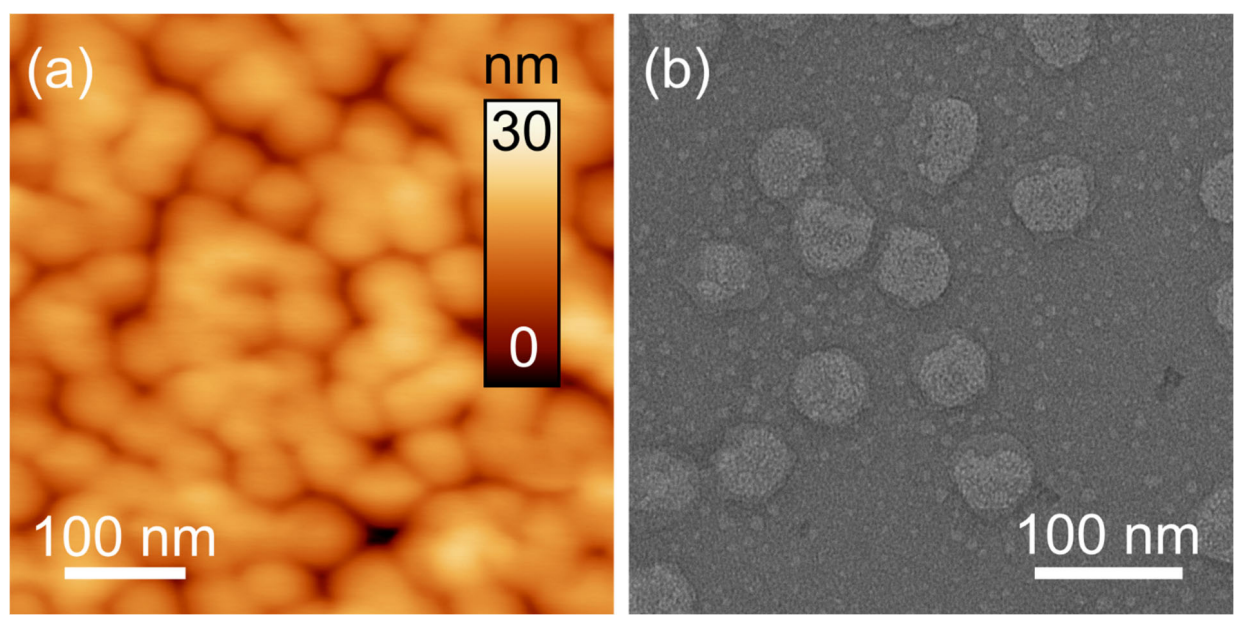

**Figure S2.** (a) High-resolution AFM image of a single bR vesicular layer on TiN. (b) High-resolution TEM image (top view) of dispersed bR vesicles on a carbon film.

## 3. Detailed FTIR and XPS spectra

### 3.1 FTIR

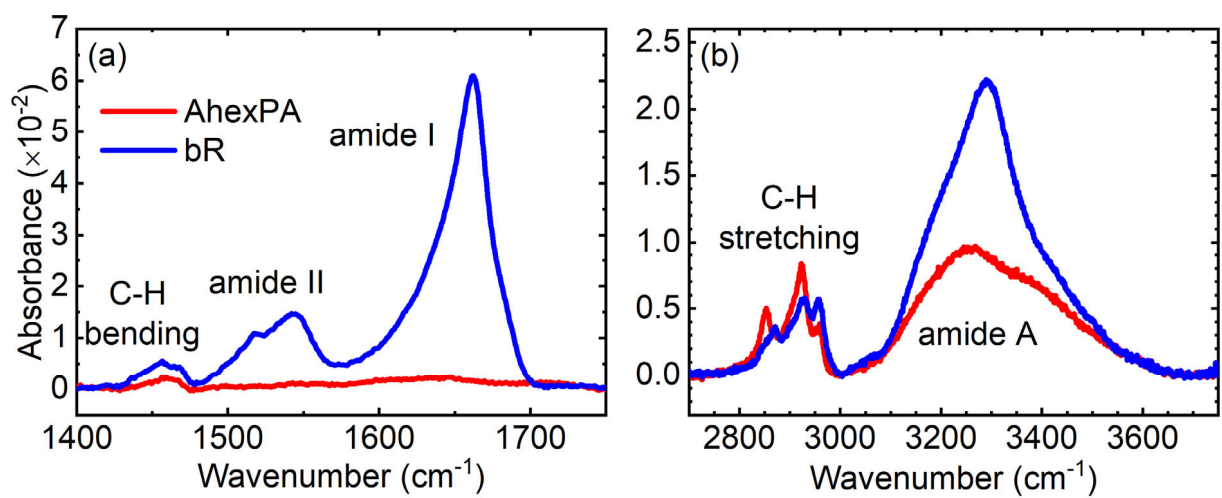

**Figure S3.** Baseline-corrected FTIR absorbance spectra of a TiN/AhexPA (red line) and a TiN/AhexPA/1bR sample (blue line) showing (a) the amide I and amide II bands, as well as the C-H bending bands and (b) the amide A (N-H stretching) and the C-H stretching bands.

**Table S1.** Summary of the characteristic FTIR absorbance bands of a TiN/AhexPA and a TiN/AhexPA/1bR sample, shown in the spectra of Figure S3.

| Band | AhexPA sample | | bR sample | |
|---|---|---|---|---|
| | Peak position [cm$^{-1}$] | Assignment | Peak position [cm$^{-1}$] | Assignment |
| N-H stretching | 3267 | amine (linker) | 3290 | amide A [a)] (protein), amine (linker) |
| C-H stretching | 2853 2923 2959 | linker backbone | 2871 2930 2957 | linker backbone, protein amino acid sequence |
| amide I [b)] | - | - | 1662 | protein |
| amide II [c)] | - | - | 1544 | protein |
| C-H bending | 1452 | linker backbone | 1456 | linker backbone, protein amino acid sequence |

a) The amide A band consists mainly of N-H stretching vibrations.; b) The amide I band consists mainly of C=O and C-N stretching vibrations.; c) The amide II band consists mainly of N-H bending, C-N stretching and C-C stretching vibrations.[17]

## 3.2 Qualitative XPS analysis

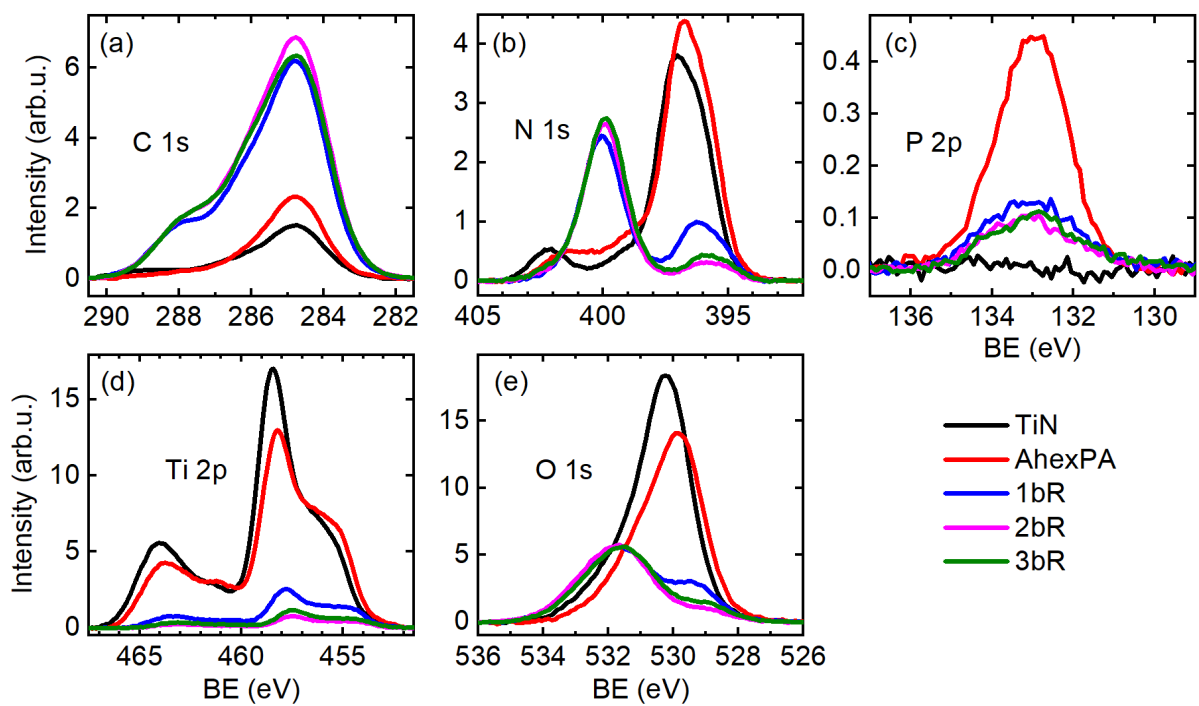

**Figure S4.** X-ray photoelectron spectra of the (a) C 1s, (b) N 1s, (c) P 2p, (d) Ti 2p, and (e) O 1s peaks for a blank TiN (black lines), a TiN/AhexPA (red lines), a TiN/AhexPA/1bR (blue lines), a TiN/AhexPA/2bR (magenta lines), and a TiN/AhexPA/3bR sample (green lines).

In the C 1s spectra of Figure S4 (a), the intensity of the C signal increases drastically after protein grafting. No significant difference is observed between protein layers of different thickness, which indicates that the thickness of the thinnest layer is already greater than the electron mean free path, assuming comparable surface coverage. The shoulder appearing only in the protein-coated samples at ~288 eV, left of the main hydrocarbon peak at 284.8 eV (charge reference), is associated with the amide moiety (peptide bond, C(=O)N),[18] and provides direct evidence of the presence of proteins in the layer. The inverse intensity trend is observed in the Ti 2p spectra (see Figure S4 (d)), as the organic layers screen the Ti signal, which originates solely from the TiN substrate. Combined, the trends observed in the C 1s and Ti 2p peak intensities, along with the detection of peptide bonds in the bR-coated samples, prove the formation of AhexPA and protein layers. Phosphorus (see Figure S4 (c)) is present in the highest concentration in the TiN/AhexPA sample and to a lesser extent in the three protein-coated samples (in similar concentration). The relatively high P signal in the TiN/AhexPA sample, combined with the Ti signal screening, proves the successful self-assembly of the AhexPA linker layer. In the bR samples, the P signal is likely the combined contribution of the underlying AhexPA layer and the phospholipids that stabilize the bR vesicles. In the N 1s spectra (Figure S4 (b)), the two most pronounced peaks appear at ~396-397 eV and ~400 eV. The one at 396-397 eV is associated with TiN[19] as it is higher in the TiN and TiN/AhexPA samples than in the protein-coated samples. The one at 400 eV is assigned to peptide bonds [20] as this peak is particularly high on the protein-coated samples. However, amino groups (NH$_2$), here present in AhexPA and the protein chain termini, possess similar binding energy. Oxygen, which is also contained in most materials (TiN/TiO$_x$, AhexPA, protein, phospholipids), displays two main spectral features (see Figure S4 (e)). The first, located at ~530 eV, may be assigned to TiO$_x$ from the substrate,[21] as these peak intensities follow the same intensity trend with the organic layer number as the Ti 2p peak intensities. The second, which is located at ~531.5 eV and is most pronounced in the protein-coated samples, is assigned to the C=O moiety[22] present in the peptide bonds of bR, but also in the residual carboxyl groups (COOH) of bR and the ester group (COO) of the phospholipids. The atomic concentrations calculated from the spectra of Figure S4 are presented in Table S2.

**Table S2.** Atomic concentrations (percent) calculated from the peak areas of the photoelectron spectra in Figure S4. Estimated relative error: 10%. Percentages may not add up to 100% due to rounding errors and trace amounts of other elements.

| Sample | C [%] | Ti [%] | N [%] | O [%] | P [%] | S [%] |
|---|---|---|---|---|---|---|
| TiN | 14 | 26 | 19 | 40 | 0 | 0 |
| AhexPA | 19 | 23 | 22 | 33 | 3 | 0 |
| 1bR | 62 | 4 | 13 | 19 | 1 | 0 |
| 2bR | 70 | 1 | 11 | 16 | 1 | 0 |
| 3bR | 68 | 2 | 12 | 17 | 1 | 0 |

### 3.3 Detailed calculation of protein layer coverage

The total protein multilayer coverage is calculated based on the screening of the Ti 2p signal intensity by the overlying organic layers (see Figure S4 (d) and Table S2, third column). Here, the surface coverage $\theta$ of the bR multilayer is defined as the ratio

$$\theta = \frac{A}{A_{tot}}, \tag{S13}$$

where $A$ is the sample area covered by protein vesicles (regardless of stack height) and $A_{tot}$ is the total sample area. The surface coverage is estimated from the XPS data using the formula

$$\theta \approx \frac{c(\text{Ti})}{c_{\text{ref}}(\text{Ti})}, \tag{S14}$$

where $c(\text{Ti})$ and $c_{\text{ref}}(\text{Ti})$ are the atomic concentrations of Ti in the protein-coated sample and in the reference (only linker-coated) sample, respectively. The Ti atomic concentrations are derived from the integrated Ti 2p peaks (Figure S4 (d)) and are shown in the third column of Table S2.

In principle, the coverage calculated with Equation (S14) may deviate from the true value given by Equation (S13), because of photoelectron transmission through the protein vesicles (imperfect screening). In the following, we will review the accuracy of Equation (S14). The photoelectron attenuation length of bR is approximated using the empirical formula

$$\lambda = 8.37 \cdot 10^{-3} \text{ nm} \left(\frac{E}{\text{eV}}\right)^{0.842}, \tag{S15}$$

derived for various organic materials,[23] where $E$ is the photoelectron kinetic energy. The calculation for Ti 2p photoelectrons ($E \approx 795.6$ eV) yields $\lambda = 2.3$ nm, implying that a single vesicle layer (net average thickness 7.4 nm by AFM) transmits only 4% of the incoming photoelectrons, whereas for thicker layers this percentage further drops exponentially. Consequently, in a bR vesicle monolayer sample, Equation (S14) underestimates the surface coverage $\theta$ by 4% on average. In the case of a bR vesicle trilayer sample (net average thickness 15.5 nm), the underestimation drops to 1%. Therefore, we use Equation (S14) for calculating the coverage of all protein-coated samples without correction.

### 3.4 Detailed calculation of vesicle protein content

The foundation for the calculation of the vesicle protein content is the peak at 400 eV in the N 1s photoelectron spectrum of the protein-coated samples (1bR-3bR, see Figure S4 (b)), which is assigned to the amide moiety (C(=O)−N, literature value: 399.7 eV [20]). Let $c(\text{N, amide})$ be the atomic concentration of N atoms in amide groups in the sample, which can be determined by deconvolution of the N 1s spectrum. Amide groups originate exclusively from the bR molecules. The elemental composition of bR is given in **Table S3**.

**Table S3**: Elemental composition of bR (source: Protein Data Bank (PDB), code 1R2N).

| Molecule | Residues | Atoms | | | | | |
|---|---|---|---|---|---|---|---|
| | | C | H | N | O | S | Total |
| bR | 232[a] | 1202 | 1854 | 275 | 306 | 9 | 3646 |
| retinal | 1 | 20 | 28 | 0 | 0 | 0 | 48 |
| water | 4 | 0 | 8 | 0 | 4 | 0 | 12 |
| total | | 237 | 1222 | 1890 | 275 | 310 | 9 | 3706 |

[a] Only modelled bR residues.

The atomic concentration $c(\text{bR})$ of atoms belonging to bR in the sample is given by

$$c(\text{bR}) = c(\text{N, bR}) + c(\text{C, bR}) + c(\text{O, bR}) + c(\text{S, bR}), \tag{S16}$$

where $c(\text{X, bR})$ is the atomic concentration of element X atoms belonging to bR in the sample. Hydrogen is neglected because it cannot be detected in XPS. $c(\text{N, bR})$ is given by

$$c(\text{N, bR}) = c(\text{N, amide}) \cdot \frac{N(\text{N,bR})}{N(\text{N,amide})} \tag{S17}$$

where $N(\text{N, bR}) = 275$ is the total number of N atoms in the chemical structure of bR (see Table S3) and $N(\text{N, amide}) = L - 1 = 231$ is the number of N atoms in amide groups in bR, with $L$ being the modelled protein amino acid sequence length, according to PDB entry 1R2N. The atomic concentrations $c(\text{C, bR})$, $c(\text{O, bR})$, and $c(\text{S, bR})$ are given by

$$c(\text{C, bR}) = c(\text{N, bR}) \cdot \frac{N(\text{C,bR})}{N(\text{N,bR})} \tag{S18}$$

$$c(\text{O}, \text{bR}) = c(\text{N}, \text{bR}) \cdot \frac{N(\text{O,bR})}{N(\text{N,bR})} \tag{S19}$$

$$c(\text{S}, \text{bR}) = c(\text{N}, \text{bR}) \cdot \frac{N(\text{S,bR})}{N(\text{N,bR})} \tag{S20}$$

$N(\text{C}, \text{bR}) = 1250$, $N(\text{O}, \text{bR}) = 310$, and $N(\text{S}, \text{bR}) = 9$ is the total number of C, O, and S atoms in the structure of bR, respectively (see Table S3), where the retinal and structural water have been included in the atom count. By substituting Equation (S17)-(S20) into Equation (S16), we obtain:

$$c(\text{bR}) = c(\text{N}, \text{amide}) \cdot \frac{N(\text{N,bR}) + N(\text{C,bR}) + N(\text{O,bR}) + N(\text{S,bR})}{N(\text{N,amide})} \tag{S21}$$

The above formula yields the protein content of the sample within the analysis volume of XPS. To improve the accuracy of the above result, we correct Equation (S21) with respect to (a) the adventitious carbon and (b) the TiN/TiO$_x$/AhexPA substrate contribution.

(a) The atomic concentration of adventitious C, denoted by $c(\text{C}, \text{adv.})$ is extracted from the spectrum of the TiN reference sample (see first row of **Table S2**). We assume that all samples carry the same concentration of adventitious C, regardless of surface coating.

(b) The contribution of the substrate, including the AhexPA linker layer, denoted by $c(\text{sub.})$, is estimated from atomic concentrations in the AhexPA linker sample (see second row of Table S2) and from the ratio $c(\text{Ti}, \text{prot.})/c(\text{Ti}, \text{sub.})$ between the Ti atomic concentrations in the protein sample and in the AhexPA reference sample (see 1$^{\text{st}}$ column of Table S2). We thus derive the following formula for the substrate contribution:

$$c(\text{sub.}) = \frac{c(\text{Ti,prot.})}{c(\text{Ti,sub.})} \cdot (c(\text{C}, \text{sub.}) - c(\text{C}, \text{adv.}) + c(\text{Ti}, \text{sub.}) + c(\text{O}, \text{sub.}) + c(\text{N}, \text{sub.}) + c(\text{P}, \text{sub.})) \tag{S22}$$

where $c(\text{X}, \text{sub.})$ are the atomic concentrations of element X in the TiN/TiO$_x$/AhexPA substrate (see second row of Table S2).

Correcting Equation (S21) with respect to the adventitious carbon and the substrate contribution, we obtain

$$c_{\text{corr}}(\text{bR}) = \frac{c(\text{bR})}{1 - c(\text{C,adv.}) - c(\text{sub.})} \tag{S23}$$

or, expressed percent,

$$c_{\text{corr}}\%(\text{bR}) = 100\% \cdot \frac{c(\text{bR})}{1-c(\text{C,adv.})-c(\text{sub.})} \quad (S24)$$

Equation (S24) was applied to all protein samples (1bR-3bR) to yield the percent protein content by atom count.

To calculate the protein content by mass, all atomic concentrations entering the calculations leading to Equation (S23) were weighted by the relative atomic mass $A_r$ of the respective element. We thus obtain

$$c_{\text{m,corr}}(\text{bR}) = \frac{c_m(\text{bR})}{m}, \quad (S25)$$

where

$$c_m(\text{bR}) = c(\text{N, amide}) \cdot \frac{N(\text{N,bR}) \cdot A_r(\text{N}) + N(\text{C,bR}) \cdot A_r(\text{C}) + N(\text{O,bR}) \cdot A_r(\text{O}) + N(\text{S,bR}) \cdot A_r(\text{S})}{N(\text{N,amide})} \quad (S26)$$

and

$$m = \sum_{X=C,N,O,P,S} \left[ \left( c(X, \text{prot.}) - \frac{c(\text{Ti,prot.})}{c(\text{Ti,sub.})} \cdot c(X, \text{sub.}) \right) \cdot A_r(X) \right] - c(C, \text{adv.}) \cdot A_r(C) \quad (S27)$$

or, expressed percent,

$$c_{\text{m,corr}}\%(\text{bR}) = 100\% \frac{c_m(\text{bR})}{m} \quad (S28)$$

Equation (S28) was applied to all protein samples (1bR-3bR) to yield the percent protein content by mass. Errors were calculated assuming a 10% relative error in all atomic concentrations.

## 4. Current-voltage measurements

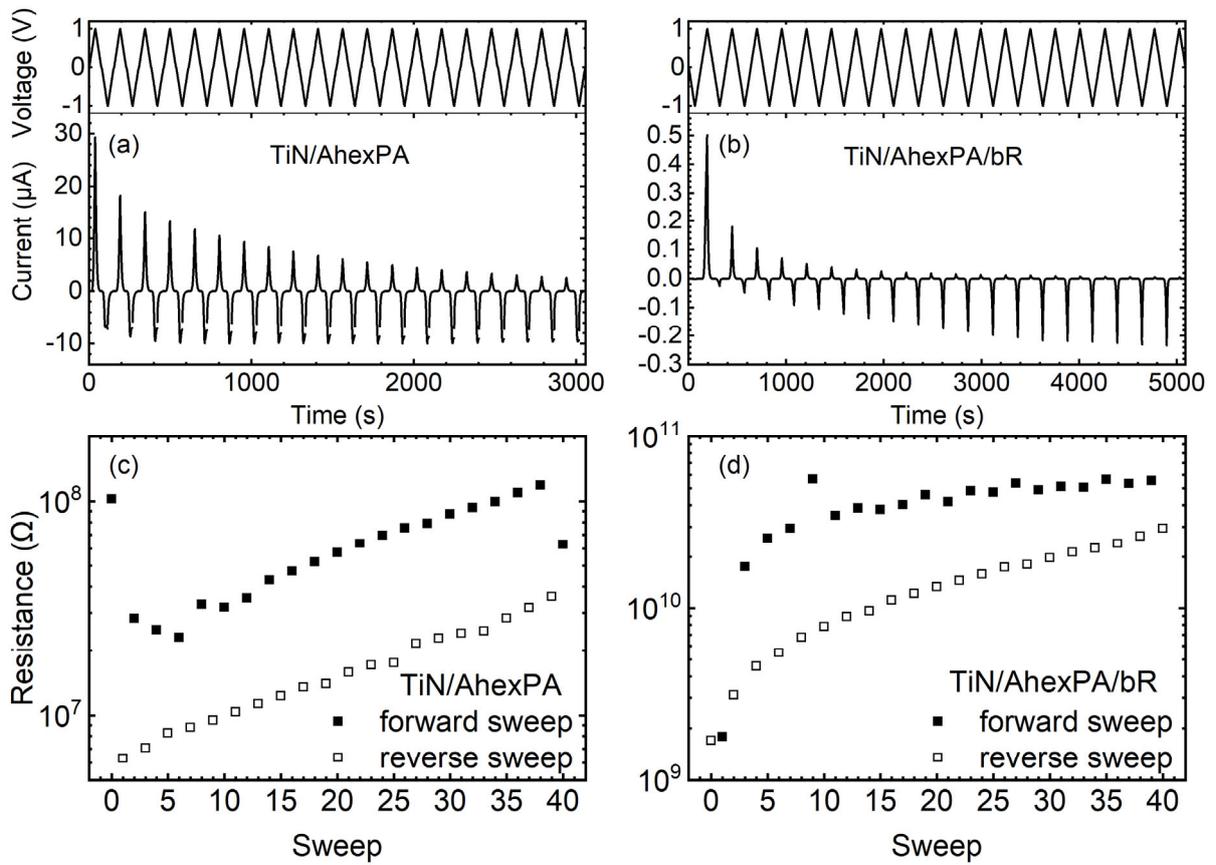

**Figure S5.** (a), (b) Time-dependence of the voltage and the current in a TiN/AhexPA/EGaIn (a) and a TiN/AhexPA/bR/EGaIn junction (b) during 20 voltage cycles between −1.0 V and +1.0 V. (c), (d) Zero-bias resistance extracted between −50 mV and +50 mV from the traces in (a) and (b), respectively, as a function of the sweep number. One cycle consists of two sweeps (forward: closed squares; reverse: open squares).

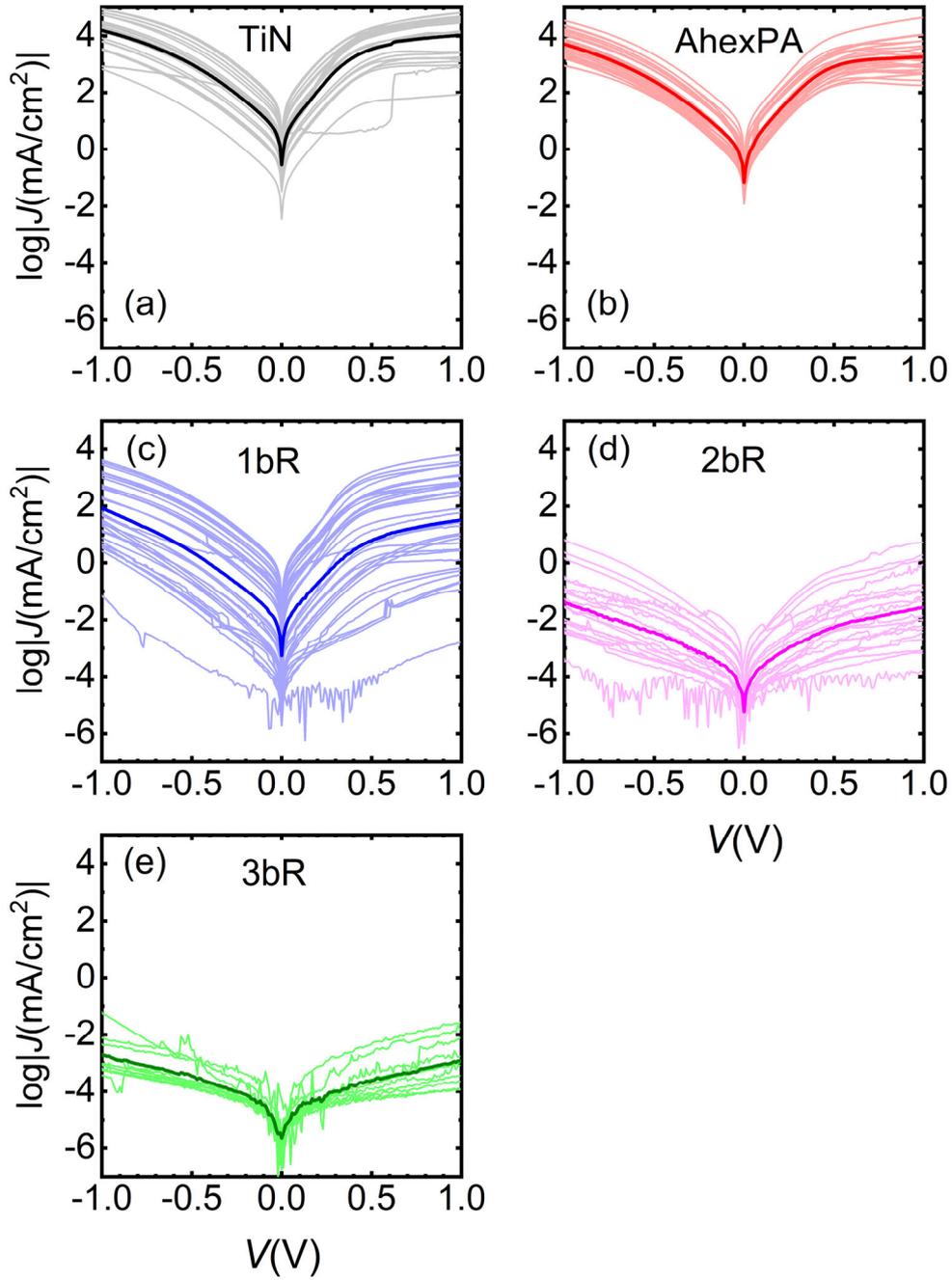

**Figure S6.** All log|*J*|-*V* traces (first sweep) from junctions with EGaIn top contacts analyzed in this work (lines in light shade), along with the mean curve for each junction type (lines in dark shade). Junction types: (a) TiN, (b) TiN/AhexPA, (c) TiN/AhexPA/1bR, (d) TiN/AhexPA/2bR, and (e) TiN/AhexPA/3bR.

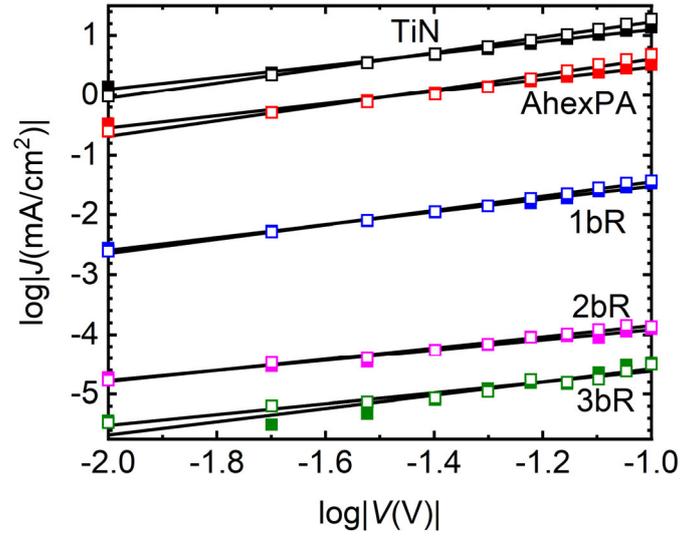

**Figure S7.** Current density-voltage characteristics of TiN (black), TiN/AhexPA (red), TiN/AhexPA/1bR (blue), TiN/AhexPA/2bR (magenta) and TiN/AhexPA/3bR (green) junctions at low bias ($|V| \leq 100$ mV), contacted with EGaIn from above at room temperature, in double-logarithmic representation (data as in Figure 4(b)). Closed squares denote the negative-bias branch and open squares denote the positive-bias branch. Black lines represent linear fits.

**Table S4.** Slopes of the linear fits to the low-bias $\log|J|$-$\log|V|$ data of **Figure** S7, for different junctions. A slope of 1 implies linear (ohmic) $J$-$V$ characteristics.

| Sample | Negative bias | | Positive bias | |
|---|---|---|---|---|
| | Slope | Error | Slope | Error |
| TiN | 1.00 | 0.03 | 1.28 | 0.03 |
| AhexPA | 1.03 | 0.04 | 1.30 | 0.07 |
| 1bR | 1.05 | 0.03 | 1.18 | 0.03 |
| 2bR | 0.86 | 0.05 | 0.94 | 0.04 |
| 3bR | 1.10 | 0.14 | 0.90 | 0.07 |

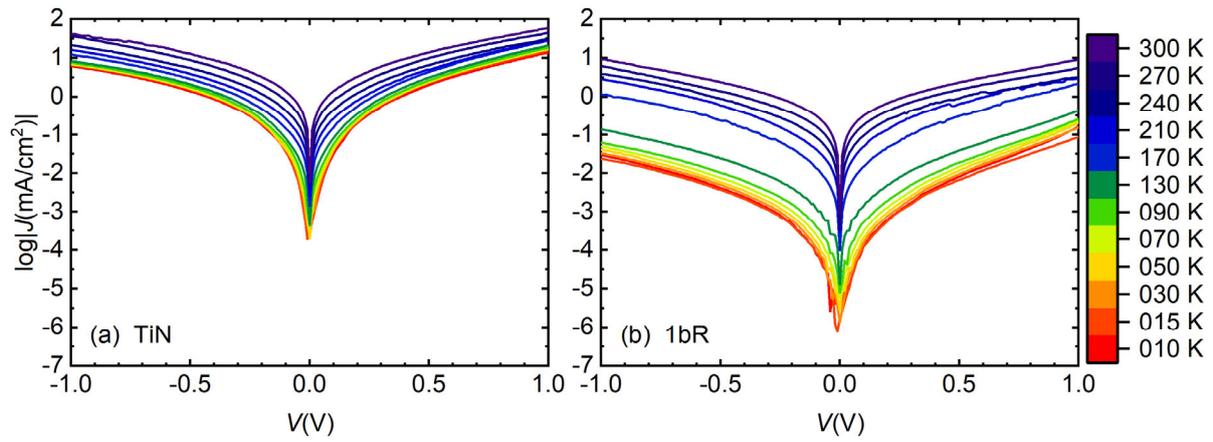

**Figure S8**: Mean log|*J*|-*V* traces (first sweep) at different temperatures (a) for TiN/Ti/Au and (b) for TiN/AhexPA/1bR/Ti/Au junctions with permanent evaporated top contacts, measured inside a cryo-probestation. The curves shown for each temperature and junction material are averaged over the same five junctions.

## 5. Voltage dependence of current attenuation coefficient

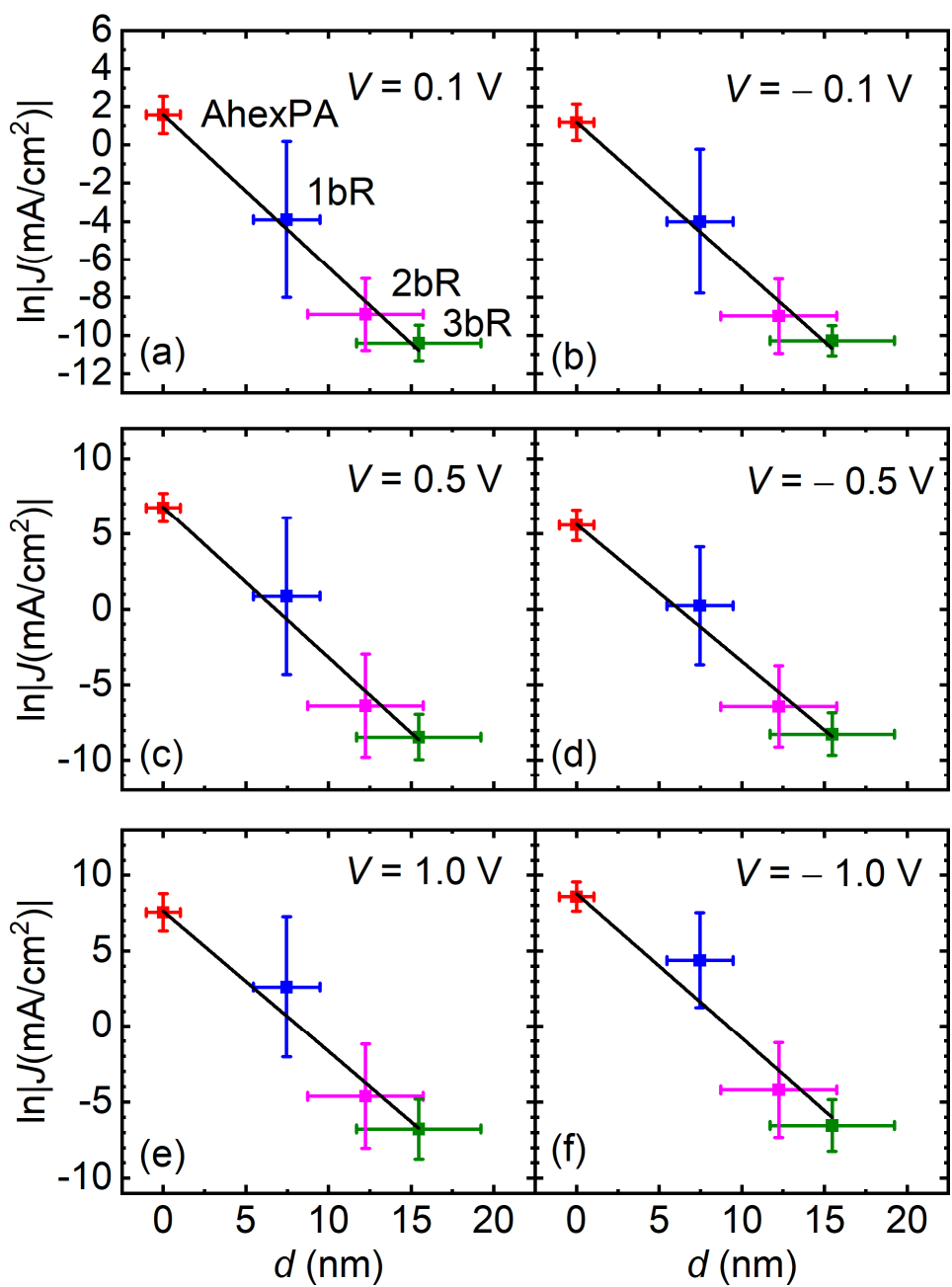

**Figure S9.** Mean logarithmic current density $\ln|J|$ as a function of the protein layer thickness $d$ for the voltage values $\pm 0.1$ V, $\pm 0.5$ V, and $\pm 1.0$ V, as indicated inside graphs (a)-(f). The surface of the AhexPA linker layer is set to $d = 0$. Squares denote the experimental data and error bars denote their standard deviation. Different point colors denote different sample types (see labels in (a)). The black lines are linear fits to the colored points. The negative of the fitted slope equals the current attenuation coefficient $\beta$.

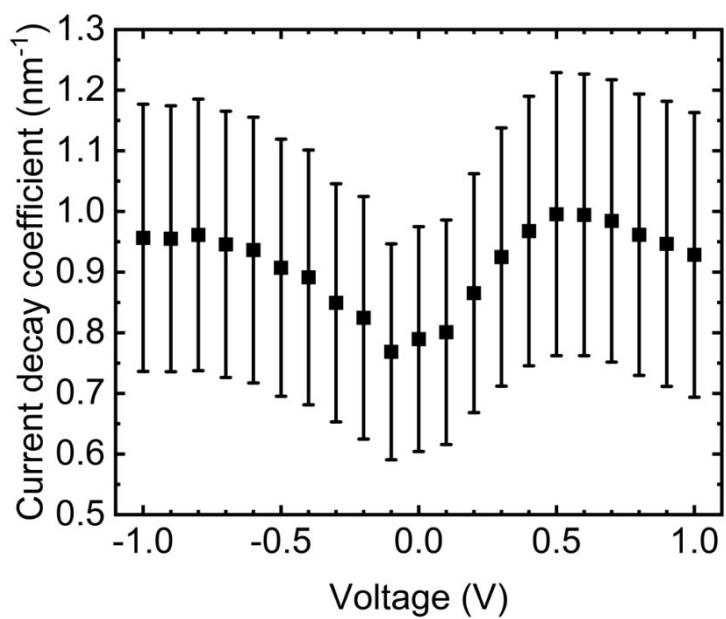

**Figure S10.** Current attenuation coefficient $\beta$ as a function of voltage $V$, extracted from the slope of the linear fit to the $\ln|J(V)|$ vs. $d$ data for $V = -1.0$ V, $-0.9$ V, …, $+1.0$ V (cf. Figure S9). At $V = 0$ V, the conductance $G$ was used instead of the current density (cf. Figure 5, main text).

## 6. Double-logarithmic conductance-length plot

In order to examine whether carrier hopping could be a possible charge transport mechanism in bR, we investigated whether the conductance $G$ measured with EGaIn top contacts as a function of protein layer thickness $d$ obeys a $1/d$ law. In the double-logarithmic plot of Figure S11, the fitted slope corresponding to the exponent of $d$ is roughly −9, instead of −1, which would be predicted in the hopping model. Therefore, carrier hopping is unlikely for the bR junctions of this study.

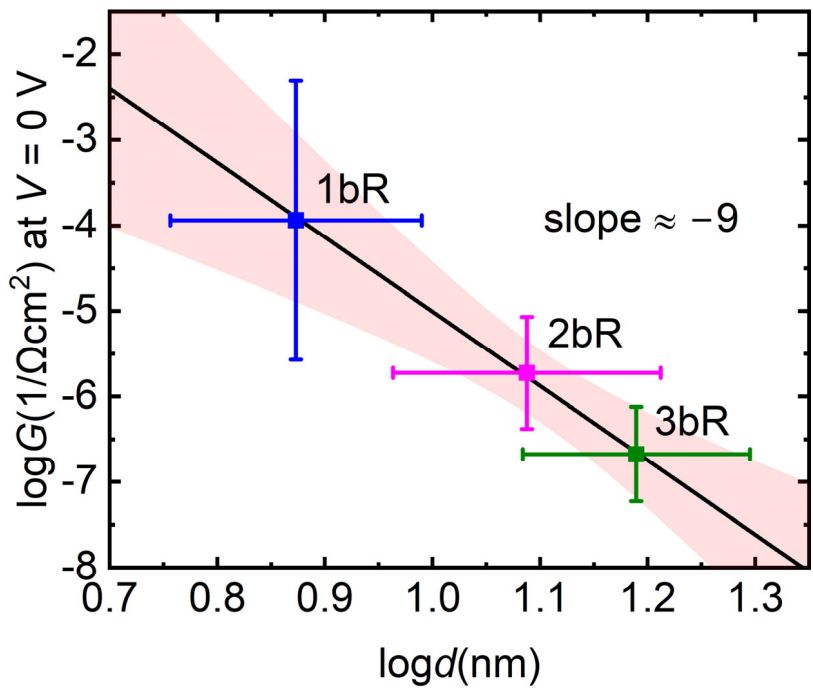

**Figure S11.** Double-logarithmic plot of the zero-bias conductance $G$ of TiN/AhexPA/1bR (blue square), TiN/AhexPA/2bR (magenta square) and TiN/AhexPA/3bR (green square) junctions contacted with EGaIn from above at room temperature, plotted against the protein layer thickness $d$, measured with AFM. The surface of the AhexPA linker layer is set to $d = 0$. The black line represents a linear fit through the experimental data points. The pink band denotes the 95% confidence interval. The slope of the fitted line is approximately –9, i.e., $G \propto d^{-9}$.

## 7. Variable-range hopping model fitting

**Table S5.** Zero-temperature current density, energy barrier, and coefficient of determination $R^2$ from fits to the ln$J$(0.1 V) vs. 1000/$T$ data from a TiN/AhexPA/bR/Ti/Au junction (see Figure S12) using five different charge transport models: temperature-dependent tunneling (Equation (3), main text), thermally-activated hopping (Equation (4)), and Mott variable-range (VR) hopping with exponent parameters $D$ = 1, 2, 3 (Equation (5)).

| Model | $J_{0K}$ | ± | $\phi$ | ± | $R^2$ |
| --- | --- | --- | --- | --- | --- |
| | [$10^4$ mA/cm$^2$] | [$10^4$ mA/cm$^2$] | [eV] | [eV] | |
| Tunneling | 2.0 | 0.3 | 0.12 | 0.01 | 0.99 |
| Hopping | 2.0 | 0.3 | 0.10 | 0.01 | 0.99 |
| VR hopping ($D$ = 1) | 1.8 | 0.3 | 2.3 | 0.4 | 0.99 |
| VR hopping ($D$ = 2) | 1.7 | 0.3 | 89 | 18 | 0.99 |
| VR hopping ($D$ = 3) | 1.7 | 0.3 | 4900 | 1300 | 0.99 |

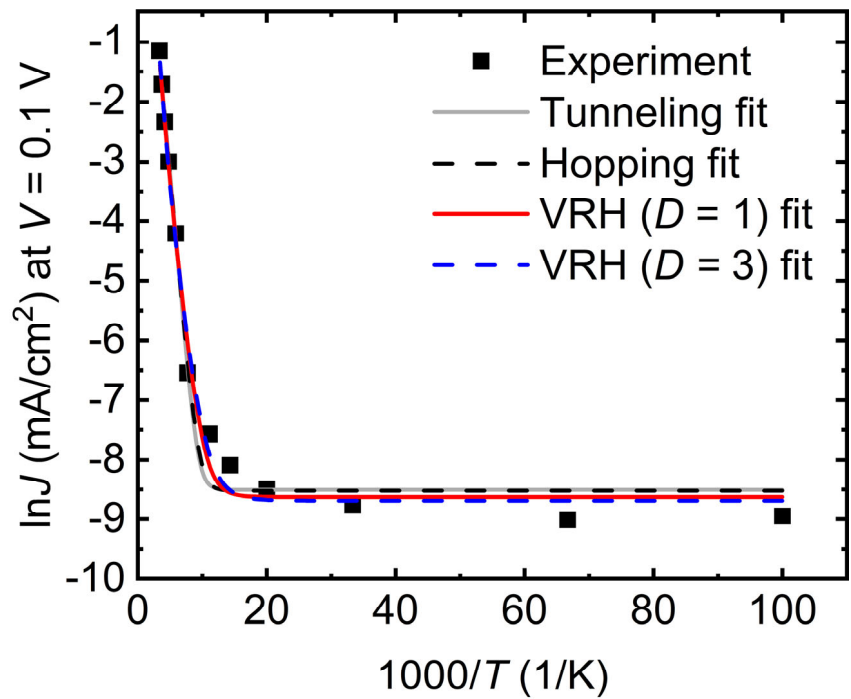

**Figure S12.** Mean ln$J$ plotted against 1000/$T$ for a TiN/AhexPA/bR/Ti/Au junction. The points represent the experimental data (same data as in Figure 6(a)). The lines represent fits to the data using tunneling (solid gray), hopping (dashed black), and Mott variable-range hopping with $D$ = 1 and 3 (solid red and dashed blue, respectively) charge transport models. The fitted curve

for Mott variable-range hopping with $D = 2$ (not shown for clarity) lies between the curves for $D = 1$ and $D = 3$.